\documentclass[nofootinbib,aps,amssymb,floatfix,superscriptaddress,preprintnumbers]{revtex4}

\usepackage[T1]{fontenc}
\usepackage{blindtext}

\usepackage{graphics,graphicx}% Include figure files
\usepackage{dcolumn}% Align table columns on decimal point
\usepackage{bm}% bold math
\usepackage{epsfig}
\usepackage[usenames]{color}
\usepackage[hidelinks]{hyperref} 
\usepackage{mathbbol}
\usepackage{epstopdf}
\usepackage{simplewick}
\usepackage[utf8]{inputenc} 
\usepackage{slashed}
\usepackage{soul}
\usepackage[dvipsnames]{xcolor}
\usepackage[export]{adjustbox}
\usepackage{amsmath}

\def\kT{k_T}

\def\Pperp{P_\perp}

\def\avkT{\la \kT^2 \ra}
\def\avpperp{\la \Pperp^2 \ra}

\newcommand{\la}{\langle}
\newcommand{\ra}{\rangle}

\begin{document}

\preprint{JLAB-THY-22-3604}

\title{{\bf Updated QCD global analysis of single transverse-spin asymmetries I: \\
Extracting $\boldsymbol{\tilde{H}}$, and the role of the Soffer bound and lattice QCD}}

\newcommand*{\PSU}{Division of Science, Penn State University Berks, Reading, Pennsylvania 19610, USA}\affiliation{\PSU}
\newcommand*{\LVC}{Department of Physics, Lebanon Valley College, Annville, Pennsylvania 17003, USA}\affiliation{\LVC}
\newcommand*{\TU}{Department of Physics,  Temple University,  Philadelphia,  PA 19122, USA}\affiliation{\TU}
\newcommand*{\JLAB}{Theory Center, Jefferson Lab, Newport News, VA 23606, USA}
\author{Leonard Gamberg}\email{lpg10@psu.edu}\affiliation{\PSU}
\author{Michel Malda}\email{mjm016@lvc.edu}\affiliation{\LVC}
\author{Joshua~A.~Miller}\email{joshua.miller0007@temple.edu}\affiliation{\TU}\affiliation{\LVC}
\author{Daniel Pitonyak}\email{pitonyak@lvc.edu}\affiliation{\LVC}
\author{Alexei Prokudin}\email{prokudin@jlab.org}\affiliation{\PSU}\affiliation{\JLAB}
\author{Nobuo Sato}\email{nsato@jlab.org\\}\affiliation{\JLAB}
\collaboration{Jefferson Lab Angular Momentum (JAM) Collaboration}

\begin{abstract}
\noindent 
We present an update  to the QCD global analysis of single transverse-spin asymmetries presented in Ref.~\cite{Cammarota:2020qcw} ({\tt JAM3D-20}).  {\tt JAM3D-20} simultaneously included transverse momentum dependent and collinear twist-3 observables, both of which are sensitive to quark-gluon-quark correlations in hadrons. 
In this study we extract for the first time the twist-3   chiral odd  fragmentation function $\tilde{H}$ by incorporating the $\sin\phi_s$ modulation data from SIDIS along with its contribution to the single transverse-spin asymmetry in pion production from proton-proton collisions. We also explore the impact of lattice QCD tensor charge calculations and the Soffer bound on our global analysis. We find that both constraints can be accommodated within our results, with $\tilde{H}$ playing a key role in maintaining agreement with the data from proton-proton collisions.

\end{abstract}

\maketitle

%%%%%%%%%%%%%%%%%%%%%%%%%%%%%%%%%%%%%%%%%%%%%%%%%%%%%%%%%%%%%%%%%%%%
\section{Introduction}
%%%%%%%%%%%%%%%%%%%%%%%%%%%%%%%%%%%%%%%%%%%%%%%%%%%%%%%%%%%%%%%%%%%%
Mapping the 3-dimensional structure of hadrons relies crucially on understanding phenomena sensitive to the transverse spin of hadrons and/or partons.  Since the late 1970s, single transverse-spin asymmetries (SSAs) have been the focus of intense experimental and theoretical efforts. These observables probe novel intrinsic parton motion and quark-gluon-quark correlations in hadrons.  From the experimental side, such measurements include $A_N$~\cite{Bunce:1976yb,Klem:1976ui,Adams:1991rw,Krueger:1998hz,Allgower:2002qi,Adams:2003fx,Adler:2005in,Lee:2007zzh,Abelev:2008af,Arsene:2008aa,Adamczyk:2012qj,Adamczyk:2012xd,Bland:2013pkt,Adare:2013ekj,Adare:2014qzo,Airapetian:2013bim,Allada:2013nsw,STAR:2020nnl} and Collins effect hadron-in-jet~\cite{STAR:2017akg,STAR:2020nnl} from proton-proton collisions, the Sivers~\cite{Airapetian:2009ae, Alekseev:2008aa,Qian:2011py,Adolph:2014zba,Zhao:2014qvx,Adolph:2016dvl,HERMES:2020ifk}, Collins~\cite{ Alekseev:2008aa,Airapetian:2010ds,Qian:2011py,Adolph:2014zba,Zhao:2014qvx,HERMES:2020ifk}, and $\sin\phi_S$~\cite{HERMES:2020ifk} asymmetries in semi-inclusive deep-inelastic scattering (SIDIS), the Collins effect in semi-inclusive electron-positron annihilation to a hadron pair (SIA)~\cite{Seidl:2008xc,TheBABAR:2013yha,Aubert:2015hha,Ablikim:2015pta,Li:2019iyt}, and the Sivers effect in Drell-Yan (DY) lepton pair~\cite{Aghasyan:2017jop} or weak gauge boson~\cite{Adamczyk:2015gyk} production. 

From a theoretical standpoint, two main frameworks have been developed to describe these data sets.  For processes with two scales $\Lambda_{QCD}\sim q_T\ll Q$, one uses transverse momentum dependent (TMD) factorization~\cite{Collins:1981uk,Collins:1981uw, Collins:1984kg,Meng:1995yn,Idilbi:2004vb,Collins:2004nx,Ji:2004xq,Ji:2004wu,Collins:2011zzd,GarciaEchevarria:2011rb} to express the cross section in terms of perturbatively calculable hard scattering coefficients and non-perturbative parton distribution and/or fragmentation functions (TMD PDFs and FFs, collectively called TMDs).  The latter depend not only on the lightcone momentum fractions carried by the partons, but also their intrinsic transverse momenta.  The TMD approach has seen considerable progress over the past decade in terms of the proper definition and evolution of TMDs (see, e.g.,~\cite{Collins:2011zzd,GarciaEchevarria:2011rb,Aybat:2011zv,Aybat:2011ge,Collins:2014jpa,Echevarria:2016scs,Gutierrez-Reyes:2018iod}) and the implementation of these rigorous details into phenomenology (see, e.g.,~\cite{Echevarria:2014xaa,Kang:2015msa,Bacchetta:2017gcc,Bacchetta:2019sam,Bertone:2019nxa,Scimemi:2019cmh,Echevarria:2020hpy,Kang:2020xgk,Bury:2021sue,Bacchetta:2020gko}). Oftentimes specific TMDs (at least at leading twist) can be isolated through a unique azimuthal modulation in the cross section\footnote{Note that recent work shows next-to-leading order collinear effects can cause additional terms to appear that may spoil this na\"{i}ve statement~\cite{Benic:2019zvg}.}. 

If an observable is only sensitive to one large scale $Q \gg \Lambda_{QCD}$, then one can employ collinear factorization, whose non-perturbative objects depend only on the lightcone momentum fractions carried by the partons.  In the case of SSAs, the collinear PDFs and FFs are subleading twist (twist-3) and encode quark-gluon-quark correlations in hadrons~\cite{Efremov:1981sh,Efremov:1984ip,Qiu:1991pp,Qiu:1991wg,Qiu:1998ia,Eguchi:2006qz,Kouvaris:2006zy,Eguchi:2006mc,Koike:2009ge, Kang:2011hk, Metz:2012ct, Beppu:2013uda}. The collinear twist-3 (CT3) framework involves a rich set of PDFs and FFs that cannot be easily isolated like TMDs.  Nevertheless, there has been successful phenomenological work explaining SSAs within the CT3 formalism~\cite{Kouvaris:2006zy,Kanazawa:2010au,Metz:2012ui,Gamberg:2013kla,Kanazawa:2014dca,Gamberg:2014eia,Gamberg:2017gle,Cammarota:2020qcw}. 

These two frameworks do not exist in isolation from each other and have been shown in previous theoretical calculations to agree in their overlapping region of validity $\Lambda_{QCD}\ll  q_T\ll Q$~\cite{Ji:2006ub,Ji:2006br,Koike:2007dg,Zhou:2008fb,Yuan:2009dw,Zhou:2009jm}.  Moreover, in Ref.~\cite{Cammarota:2020qcw} ({\tt JAM3D-20}) the authors demonstrated for the first time that one can perform a simultaneous QCD global analysis of SSAs in SIDIS (Sivers and Collins effects), SIA (Collins effect), DY (Sivers effect), and proton-proton collisions ($A_N$ for pion production) and extract a universal set of non-perturbative functions. The work in Ref.~\cite{Cammarota:2020qcw} was based on a parton model analysis of the relevant SSAs (see Secs.~\ref{s:Theory}, \ref{s:Method} for more details). Adopting the factorization hypothesis of this framework, it was demonstrated that one can describe the experimental data on SSAs and obtain, for the first time, agreement with lattice QCD on the values of the nucleon tensor charges.  The results of Ref.~\cite{Cammarota:2020qcw} significantly bolstered the claim that all SSAs have a common origin.

Several extensions to the {\tt JAM3D-20} analysis allow for further tests of the conclusions of Ref.~\cite{Cammarota:2020qcw}.  For example, one can incorporate other SSAs into the framework that were not in {\tt JAM3D-20} (such as Collins effect hadron-in-jet~\cite{STAR:2017akg,STAR:2020nnl} or $A_N$ for jets~\cite{Bland:2013pkt,STAR:2020nnl}); likewise, new data from observables that were already a part of {\tt JAM3D-20} can be included. In addition, one can explore the impact of theoretical constraints, like lattice QCD tensor charge calculations and the Soffer bound on transversity.  In order to clearly study the effects all of these have on the {\tt JAM3D-20} results, and more generally on our understanding of the mechanism underlying SSAs, we separate our updated analysis into two parts.  In this paper we extract the quark-gluon-quark FF $\tilde{H}(z)$ and study the role lattice QCD data on the nucleon tensor charge~\cite{Gupta:2018qil,Yamanaka:2018uud,Hasan:2019noy,Alexandrou:2019brg} and the Soffer bound on transversity~\cite{Soffer:1994ww} play in our simultaneous global analysis.  

{\tt JAM3D-20} (and also Ref.~\cite{Gamberg:2017gle}) did not include the contribution of $\tilde{H}(z)$ to $A_N$ due to the lack of SIDIS data directly sensitive to it and the inability of other data sets to constrain the function.  However, with the published measurements from HERMES on the $A_{UT}^{\sin\phi_S}$ asymmetry in SIDIS now available~\cite{HERMES:2020ifk}, we are in a position to obtain the first information on $\tilde{H}(z)$ within a global analysis; thereby we can include all pieces from the theoretical calculation of the $A_N$ fragmentation (Collins-type) term that drives the asymmetry~\cite{Kanazawa:2014dca,Gamberg:2017gle,Cammarota:2020qcw}.  This allows us to then more completely understand how lattice QCD~\cite{Gupta:2018qil,Yamanaka:2018uud,Hasan:2019noy,Alexandrou:2019brg}  and the Soffer bound~\cite{Soffer:1994ww} influence our extracted non-perturbative functions. 

The paper is organized as follows:~in Sec.~\ref{s:Theory} we discuss the theoretical setup for all observables used in our global analysis. The new  results are presented in Sec.~\ref{s:Pheno}, which is subdivided into an overview of our methodology in Sec.~\ref{s:Method}, the non-perturbative functions we extracted in Sec.~\ref{s:Results}, and a comparison of theory with experimental data in Sec.~\ref{s:Discuss}.  An exploratory study on the role of antiquarks is given in Sec.~\ref{s:Antiquarks}.
Conclusions and an outlook are presented in Sec.~\ref{s:Concl}.

%%%%%%%%%%%%%%%%%%%%%%%%%%%%%%%%%%%%%%%%%%%%%%%%%%%%%%%%%%%%%%%%%%%%
\section{Theoretical Background} \label{s:Theory}
%%%%%%%%%%%%%%%%%%%%%%%%%%%%%%%%%%%%%%%%%%%%%%%%%%%%%%%%%%%%%%%%%%%%
The {\tt JAM3D-20} global analysis included the Sivers $A_{UT}^{\sin(\phi_h-\phi_S)}$ and Collins $A_{UT}^{\sin(\phi_h+\phi_S)}$ asymmetries in SIDIS, Collins asymmetry in SIA for so-called unlike-like ($A_{UL}$) and unlike-charged ($A_{UC}$) ratios, Sivers asymmetry in DY for $W^\pm\!/Z$ production ($A_N^{W/Z}$) and for
$\mu^+\mu^-$ production ($A_{T,\mu^+\mu^-}^{\sin\phi_S}$), and $A_N$ for pion production in proton-proton collisions ($A_N^\pi$).  In the following we review our parton model framework for these observables.  The versions needed for the TMD evolution/Collins-Soper-Sterman formulation of SIDIS, SIA, and DY can be found, e.g., in Refs.~\cite{Echevarria:2020hpy,Kang:2015msa}.  We also discuss the theory for the $A_{UT}^{\sin\phi_S}$ asymmetry in SIDIS (Sec.~\ref{s:SIDIS}) as well as expressions for the nucleon tensor charges and Soffer bound on transversity (Sec.~\ref{s:gT_SB}), which are all new pieces to this analysis.

%%%%%%%%%%%%%%%%%%%%%%%%%%%%%%%%
%%%%%%%%%%%%%%%%%%%%%%%%%%%%%%%%
\subsection{Semi-inclusive deep-inelastic scattering \label{s:SIDIS}}
%%%%%%%%%%%%%%%%%%%%%%%%%%%%%%%%
%%%%%%%%%%%%%%%%%%%%%%%%%%%%%%%%

The SIDIS reaction involving an unpolarized lepton scattering on a transversely polarized nucleon,
\begin{equation}
    \ell(l) + N(P,S_T) \rightarrow \ell(l') + h(P_h) + X\,,
\end{equation}
has several SSAs that can be studied.  We focus on the Sivers effect $A_{UT}^{\sin(\phi_h-\phi_S)}$, Collins effect~$A_{UT}^{\sin(\phi_h+\phi_S)}$, and $\sin\phi_S$ asymmetry $A_{UT}^{\sin\phi_S}$. 
These asymmetries can be expressed as ratios of structure functions~\cite{Bacchetta:2006tn}:
\begin{equation} \label{e:SIDIS_SSAs}
    A_{UT}^{\sin(\phi_h-\phi_s)} = \frac{F^{\sin(\phi_h-\phi_s)}_{UT}}{F_{UU}}\,, \quad\quad A_{UT}^{\sin(\phi_h+\phi_s)} = \frac{F^{\sin(\phi_h+\phi_s)}_{UT}}{F_{UU}}\,, \quad\quad \left\langle \! A_{UT}^{\sin\phi_s}\!\right\rangle_{\!\!P_{hT}} = \frac{\int\! d^2\!\vec{P}_{hT}\, F^{\sin\phi_s}_{UT}}{\int \!d^2\!\vec{P}_{hT}\, F_{UU}}\,,
\end{equation}
where $\phi_{S}$ is the azimuthal angle of the (transverse) spin vector $S_T$ of the nucleon, and $\phi_{h}$ is the azimuthal angle of the transverse momentum $\vec{P}_{hT}$ of the produced hadron, both relative to the leptonic plane (plane formed by the incoming and outgoing leptons).  The angled brackets around $A_{UT}^{\sin\phi_S}$ indicate that we are only considering the case where the numerator and denominator are integrated over $\vec{P}_{hT}$. We note that the Collins and $\sin\phi_S$ asymmetries are sometimes written with so-called ``depolarization factors'' included.  However, COMPASS always removes these coefficients when reporting their measurements~\cite{Alekseev:2008aa,Adolph:2014zba,Adolph:2016dvl}, and HERMES in their latest publication (which supersedes earlier measurements) presented data with and without depolarization factors~\cite{HERMES:2020ifk}.  Since we opted to use the latter format, the depolarization factors are removed from Eq.~(\ref{e:SIDIS_SSAs}).

The generic structure function $F_{XY}^Z$, where $X$ is the polarization of the lepton, $Y$ the polarization of the nucleon, and $Z$ the relevant azimuthal modulation, also depends on $Q^{2} = -(l-l')^2 \equiv  -q^2$, $x_B = Q^{2}/2P \cdot q$, and $z_{h} = P \cdot P_{h}/P \cdot q$.  We work in the one-photon exchange approximation at leading order; therefore, $Q^{2}$ is the virtuality of the photon and $x_B$ ($z_h$) is equivalent to the fraction $x$ ($z$) of the incoming nucleon's (fragmenting quark's) momentum carried by the struck quark (produced hadron).  
Since we are studying the domain of small transverse momentum, the scale separation is dictated by the quantity $\vec{q}_T = -\vec{P}_{hT}/z$, which,  up to power corrections, is the transverse momentum of the exchanged photon in the frame where the incoming and outgoing hadrons are back-to-back.

The structure functions in Eq.~(\ref{e:SIDIS_SSAs}) involve convolutions of TMDs and can be written explicitly as~\cite{Bacchetta:2006tn}
\begin{align} 
    \hspace{-0.2cm}F_{UU} = \mathcal{C}_{\rm SIDIS}\!\left[f_1D_1\right], \quad\quad &F^{\sin(\phi_h-\phi_s)}_{UT} = -\mathcal{C}_{\rm SIDIS}\!\!\left[\frac{\hat{h}\cdot \vec{k}_T} {M}f_{1T}^\perp D_1\right], \nonumber\\
    & \label{e:F_convol}\\
    F^{\sin(\phi_h+\phi_s)}_{UT} = -\mathcal{C}_{\rm SIDIS}\!\!\left[\frac{\hat{h}\cdot \vec{p}_T} {M_h}h_{1} H_1^\perp\right], &\quad\quad F_{UT}^{\sin\phi_S} = \frac{2M}{Q}\mathcal{C}_{\rm SIDIS}\!\!\left[-\frac{M_h}{M}h_1\frac{\tilde{H}}{z}+\dots\right],\nonumber
\end{align} 
where $\hat{h}\equiv \vec{P}_{hT}/|\vec{P}_{hT}|$, $M$ ($M_h$)  is the nucleon (hadron) mass, and $\vec{k}_T$ ($\vec{p}_T$) is the transverse momentum of the incoming (outgoing) quark.  The convolution $\mathcal{C}_{\rm SIDIS}[wfD]$ for generic TMDs $f$ and $D$ with weight $w$ is defined as
\begin{equation}
    \mathcal{C}_{\rm SIDIS}\!\left[wfD\right] \equiv x \sum_q e_q^2\int \!d^2\vec{k}_Td^2\vec{p}_T\,\delta^{(2)}\!(\vec{k}_T-\vec{p}_T-\vec{P}_{hT}/z)\,w(\vec{k}_T,\vec{p}_T)\,f^{q/N}(x,\vec{k}_T^2)\, D^{h/q}(z,z^2\vec{p}_T^{\,2})\,,
\end{equation}
where the sum is over all light quarks and antiquarks, and $e_q$ is the charge for a specific flavor in units of the electric charge  $e$ of the positron. The functions entering Eq.~(\ref{e:F_convol}) are the unpolarized TMD PDF $f_1$ and FF $D_1$, the transversity TMD PDF $h_1$, the Sivers TMD PDF $f_{1T}^\perp$, the Collins TMD FF $H_1^\perp$, and the quark-gluon-quark TMD FF $\tilde{H}$. The  $\vec{P}_{hT}$-integrated $F_{UU}$ and $F_{UT}^{\sin\phi_S}$ structure functions (needed in the last equality of Eq.~(\ref{e:SIDIS_SSAs})) read~\cite{Bacchetta:2006tn}
\begin{equation}
    \int \!d^2\!\vec{P}_{hT}\, F_{UU} = x\sum_q e_q^2\, f_1^{q/N}(x)\,D_1^{h/q}(z)\,,\quad\quad \int \!d^2\!\vec{P}_{hT}\, F_{UT}^{\sin\phi_S} = -\frac{x}{z}\sum_q e_q^2\,\frac{2M_h} {Q} h_1^{q/N}(x)\,\tilde{H}^{h/q}(z)\,. \label{e:sinphiS}
\end{equation}
We emphasize that the (unintegrated) $F_{UT}^{\sin\phi_S}$ structure function in~(\ref{e:F_convol}) has five more terms (represented by the ellipsis) involving the coupling of various twist-2 and twist-3 TMDs.  However, upon integration over $\vec{P}_{hT}$ only one term survives ~\cite{Bacchetta:2006tn} that involves the (collinear) transversity function $h_1(x)$ and (collinear twist-3) FF $\tilde{H}(z)$.  The latter function also enters the pion $A_N$ asymmetry ($A_N^\pi$) in proton-proton collisions (see Sec.~\ref{s:AN}).  This FF is of interest because both $\tilde{H}(z)$ and the first moment of the Collins function $H_1^{\perp(1)}(z)$ can be written as an integral of the same quark-gluon-quark fragmentation correlator~\cite{Kanazawa:2015ajw}. Therefore, the Collins and ($\vec{P}_{hT}$-integrated) $\sin\phi_S$ asymmetries may have a similar underlying mechanism.  

In {\tt JAM3D-20} the only observable sensitive to $\tilde{H}(z)$ for which finalized data was available was $A_N^\pi$.  However, in $A_N^\pi$ several other terms enter that are numerically more significant than the one involving $\tilde{H}(z)$; therefore, $\tilde{H}(z)$ was set to zero in {\tt JAM3D-20} because it became a source of noise in the fit and was found to be consistent with zero. With data now available from HERMES on $A_{UT}^{\sin\phi_S}$, we have an opportunity to gain information on $\tilde{H}(z)$.  The caveat is that one must integrate over all $\vec{P}_{hT}$  in order to be directly sensitive to $\tilde{H}(z)$ (cf.~Eq.~(\ref{e:sinphiS})), which experimentally cannot be done.  For this reason, we only use the $x$- and $z$-projected $A_{UT}^{\sin\phi_s}$ HERMES data in our analysis and make the assumption that $\left\langle \! A_{UT}^{\sin\phi_s}\!\right\rangle_{\!\!P_{hT}}$ defined in (\ref{e:SIDIS_SSAs}) (and using Eq.~(\ref{e:sinphiS})) is a reasonable approximation in those cases.

%%%%%%%%%%%%%%%%%%%%%%%%%%%%%%%%
%%%%%%%%%%%%%%%%%%%%%%%%%%%%%%%%
\subsection{Semi-inclusive electron-positron annihilation to a hadron pair \label{s:epem}}
%%%%%%%%%%%%%%%%%%%%%%%%%%%%%%%%
%%%%%%%%%%%%%%%%%%%%%%%%%%%%%%%%
Let us consider the production of two almost back-to-back hadrons $h_1,h_2$ in electron-positron annihilation (SIA), mediated by a virtual photon of momentum $q$,
\begin{align}
 e^+(l) + e^-(l') \to h_1(P_{h_1}) + h_2(P_{h_2}) + X\,.\end{align}
The hadrons arise from the fragmentation of a quark or antiquark in the underlying partonic subprocess, and they carry the longitudinal momentum fractions
\begin{align}
 z_{1} = \frac{2 |\vec{P}_{h_1}|}{Q} \,,\quad z_{2} = \frac{2 |\vec{P}_{h_2}|}{Q}
\,,\end{align}
where the center-of-mass energy $(l+l')^2=q^2\equiv Q^2$.
We work in a frame where one aligns the $z$-axis along $h_2$ and measures the azimuthal angle $\phi_0$ of $h_1$ with respect to this axis and the leptonic plane (where $\vec{l}$ and $\vec{l}'$ are at a polar angle $\theta$)\footnote{There is also another frame where one defines the thrust axis and measures two azimuthal angles $\phi_1$ and $\phi_2$.  The Collins effect then manifests itself as a $\cos(\phi_1+\phi_2)$ asymmetry.  However, this asymmetry cannot be directly described within TMD factorization, so we will only consider the frame discussed in the main text (and associated data).}.  The relevant structure functions are~\cite{Boer:2008fr,Pitonyak:2013dsu},\footnote{We use $u$ in this context for ``unpolarized'' and reserve $U$ for ``unlike-sign.''}
\begin{eqnarray}
   F^{h_1h_2}_{uu} &=& \phantom{-}
   {\cal C}_{\rm SIA}[D_1 \bar D_1]\,, \nonumber \\[0.3cm]
   F^{h_1h_2}_{\cos \!2\phi_0} &=&  {\cal C}_{\rm SIA}\!\left[\frac{2(\hat{h}\cdot\vec{p}_{1T})(\hat{h}\cdot{\vec{p}}_{2T})-\vec{p}_{1T}\cdot\vec{p}_{2T}} {M_{h_1} M_{h_2}} H_1^\perp \bar H_1^\perp\right], \label{Eq:SFepems}
\end{eqnarray}
where 
\begin{align} \label{eq:def-convolution-integral-ee}
	&\hspace{-0.25cm}{\cal C}_{\rm SIA}\!\left[w\, D_1\bar D_2\right] 
	= \sum_q e_q^2 \int d^2\vec{p}_{1T} \, d^2\vec{p}_{2T}
	\; \delta^{(2)\!\!}\left( \vec{p}_{1T}+\vec{p}_{2T}  -\vec{q}_T \right)w(\vec{p}_{1T}, \vec{p}_{2T})D_1^{h_1/q}(z_{1},z_1^2\vec{p}_{1T}^{\,2})D_2^{h_2/\bar q}(z_{2},z_2^2\vec{p}_{2T}^{\,2})\,.
\end{align}
We note that $\hat{h}\equiv \vec{P}_{h\perp}/|\vec{P}_{h\perp}|$, where $\vec{P}_{h\perp}=-z_1\vec{q}_T$ is the component of $P_{h_1}$ that is transverse to $P_{h_2}$ ($\vec{q}_T$ is the transverse momentum of the virtual photon in the hadronic center-of-mass frame).
The unlike-like (UL) and unlike-charged (UC) ratios are given by~\cite{Anselmino:2007fs,Kang:2014zza,Anselmino:2015sxa},
\begin{align}
  A_{UL}(z_1,z_2,\theta,P_{h\perp}) &\equiv \frac{\langle \sin^2\theta\rangle}{\langle 1+\cos^2\theta\rangle}\left(\frac{F_{\cos \!2\phi_0}^{U}}{F^U_{uu}}-\frac{F_{\cos  \!2\phi_0}^{L}}{F^L_{uu}}\right),\label{e:UL}\\
  %%%
  A_{UC}(z_1,z_2,\theta,P_{h\perp}) &\equiv \frac{\langle \sin^2\theta\rangle}{\langle 1+\cos^2\theta\rangle}\left(\frac{F_{\cos \!2\phi_0}^{U}}{F^U_{uu}}-\frac{F_{\cos  \!2\phi_0}^{C}}{F^C_{uu}}\right),\label{e:UC}\\
  %%%
  A_{UL}(z_1,z_2,\theta) &\equiv \frac{\langle \sin^2\theta\rangle}{\langle 1+\cos^2\theta\rangle}\left(\frac{\int dP_{h\perp}P_{h\perp}F_{\cos \!2\phi_0}^{U}}{\int dP_{h\perp}P_{h\perp}F^U_{uu}}-\frac{\int dP_{h\perp}P_{h\perp}F_{\cos  \!2\phi_0}^{L}}{\int dP_{h\perp}P_{h\perp}F^L_{uu}}\right),\label{e:ULint}\\
  %%%
   A_{UC}(z_1,z_2,\theta) &\equiv \frac{\langle \sin^2\theta\rangle}{\langle 1+\cos^2\theta\rangle}\left(\frac{\int dP_{h\perp}P_{h\perp}F_{\cos \!2\phi_0}^{U}}{\int dP_{h\perp}P_{h\perp}F^U_{uu}}-\frac{\int dP_{h\perp}P_{h\perp}F_{\cos  \!2\phi_0}^{C}}{\int dP_{h\perp}P_{h\perp}F^C_{uu}}\right), \label{e:UCint}
\end{align}
where the unlike-signed~(U), like-signed~(L), and charged~(C) combinations read
\begin{equation}
F^U \equiv F^{\pi^+\pi^-}\!\!+F^{\pi^-\pi^+},\quad F^L \equiv F^{\pi^+\pi^+}\!\!+F^{\pi^-\pi^-},\quad F^C\equiv F^U\!+F^L\,.  
\end{equation}
Whether one uses (\ref{e:UL}), (\ref{e:UC}) or (\ref{e:ULint}), (\ref{e:UCint}) depends on if the measurement was differential in $P_{h\perp}$ or integrated over it.

%%%%%%%%%%%%%%%%%%%%%%%%%%%%%%%%
%%%%%%%%%%%%%%%%%%%%%%%%%%%%%%%%
\subsection{Drell-Yan lepton pair or weak gauge boson production}
%%%%%%%%%%%%%%%%%%%%%%%%%%%%%%%%
%%%%%%%%%%%%%%%%%%%%%%%%%%%%%%%%
We begin by considering pion-induced DY on a transversely polarized target with a $\mu^+\mu^-$ pair in the final state, as at COMPASS,
\begin{equation}
    \pi(P_\pi)+p(P,S_T)\to \mu^+(l)+\mu^-(l')+X\,.
\end{equation}
At LO the $\mu^+\mu^-$ pair is produced from
the annihilation of a quark and antiquark carrying the fractions $x_1$, 
$x_2$ of the longitudinal momenta of the pion and the proton, respectively. They annihilate into a virtual photon of momentum $q$, with $q^2\equiv Q^2$, and we define the unit vector $\hat{h}\equiv\vec{q}_T/|\vec{q}_T|$.
The relevant structure functions are~\cite{Arnold:2008kf} 
\begin{equation}
   F_{UU}^1 =
   {\cal C}_{\rm DY} \!\! \left[f_{1}\bar{f}_{1}\right], \quad
   F_{UT}^1 = 
   {\cal C}_{\rm DY}\!\! \left[ \frac{\hat{h} \cdot \vec{k}_{2T}}{M}
     f_1\bar{f}_{1T}^{\perp}  \right]. \\
    \label{e:DY_SFs} 
\end{equation}
The Sivers asymmetry as measured by COMPASS is then
\begin{equation}
    A_{T,\mu^+\mu^-}^{\sin\!\phi_S}=\frac{F_{UT}^{1}}{F_{UU}^1 }\,,
\end{equation}
where $\phi_S$ is the azimuthal angle of the (transverse) spin vector $S_T$.
The convolution integral in Eq.~(\ref{e:DY_SFs})
is defined as~\cite{Arnold:2008kf}
\begin{align}
 \label{Eq:convolution-integral} 
    {\cal C}_{\rm DY}&\!\!\left[w f_1 \bar{f}_2\right] = \frac{1}{N_c} 
    \sum_q e_q^2\int d^2\vec{k}_{1T} \, d^2\vec{k}_{2T} \, \delta^{(2)}\!(\vec{q}_T- \vec{k}_{1T}-\vec{k}_{2T}) \,
    w(\vec{k}_{1T},\vec{k}_{2T})f_1^{q/\pi}(x_1,\vec{k}_{1T}^2)f_2^{\bar q/p}(x_2,\vec{k}_{2T}^2)\,. %\nonumber\\ 
\end{align}  
For the case of COMPASS, a pion beam collides with a fixed-target proton, so we will use $x_{\rm beam}\equiv x_1$ and $x_{\rm target}\equiv x_2$.  Note again that $\sum_q$ runs over all light quarks and antiquarks so that, in particular, $F_{UT}^1$ in Eq.~(\ref{e:DY_SFs}) receives contributions where $f_1$ for an antiquark couples to $f_{1T}^\perp$ for a quark.

We now focus on the case of weak gauge boson production from the collision of a transversely polarized proton and unpolarized proton, as at STAR,
\begin{equation}
   p(P,S_T)+p(P')\to \{W^+,W^-,\,{\rm or}\, Z\}+X\,,
\end{equation}
where the boson carries momentum $q$ and has rapidity $y$.  The relevant structure functions are  \begin{equation}
   F_{UU}^1 =
   {\cal C}_{\rm ew} \!\! \left[f_{1}\bar{f}_{1}\right], \quad
   F_{TU}^1 = 
   -{\cal C}_{\rm ew}\!\! \left[ \frac{\hat{h} \cdot \vec{k}_{1T}}{M}
     f_{1T}^{\perp}\bar{f}_1  \right],  \\
    \label{e:WZ_SFs} 
\end{equation}
and the Sivers asymmetry as measured by STAR is then
\begin{equation}
    A_{N}^{W/Z}=\frac{F_{TU}^1}{F_{UU}^1 }\,.
\end{equation}
The convolution integral in Eq.~(\ref{e:WZ_SFs}) for a vector boson $V=W\;{\rm or}\; Z$ is defined as~\cite{Anselmino:2016uie,Echevarria:2020hpy,Bury:2021sue,Bacchetta:2020gko}
\begin{align}
 \label{Eq:convolution-integral_ew} 
    {\cal C}_{\rm ew}&\!\!\left[w f_1 \bar{f}_2\right] = \frac{1}{N_c} 
    \sum_{q_1,q_2} e_{q_1q_2,V}^2\int d^2\vec{k}_{1T} \, d^2\vec{k}_{2T} \, \delta^{(2)}\!(\vec{q}_T- \vec{k}_{1T}-\vec{k}_{2T}) \,
    w(\vec{k}_{1T},\vec{k}_{2T})f_1^{q_1/p^\uparrow\!\!}(x_1,\vec{k}_{1T}^2)f_2^{ q_2\!/p}(x_2,\vec{k}_{2T}^2) \,,
\end{align}  
where $e_{q_1q_2,W}^2=|V_{q_1q_2}|^2$ and $e_{q_1q_2,Z}=(V_{q_1}^2+A_{q_1}^2)\delta_{q_1\bar{q}_2}$, with $V_{q_1q_2}$ the elements of the CKM matrix and $V_q$ and $A_q$ the vector and axial couplings, respectively, of the $Z$ boson to a quark or antiquark of flavor $q$.  The symbol $p^\uparrow$ is used to indicate specifically which function is for the transversely polarized proton.  

We note that $f_{1T}^\perp$ written in Eqs.~(\ref{e:DY_SFs}), (\ref{e:WZ_SFs}) is the Sivers function for the DY process, which is related to the $f_{1T}^\perp$ in Eq.~(\ref{e:F_convol}) for SIDIS by~\cite{Collins:2002kn}
\begin{equation}
    f_{1T}^{\perp}(x,\vec{k}_T^2)\big |_{\rm SIDIS}= - f_{1T}^{\perp}(x,\vec{k}_T^2)\big |_{\rm DY}\,.
\end{equation}
We explicitly account for this sign change in our  analysis and extract $f_{1T}^{\perp}(x,\vec{k}_T^2) |_{\rm SIDIS}$; any plots shown of the Sivers function are for the SIDIS version. 

%%%%%%%%%%%%%%%%%%%%%%%%%%%%%%%%
%%%%%%%%%%%%%%%%%%%%%%%%%%%%%%%%
\subsection{$A_N$ for single-inclusive hadron production in proton-proton collisions} \label{s:AN}
%%%%%%%%%%%%%%%%%%%%%%%%%%%%%%%%
%%%%%%%%%%%%%%%%%%%%%%%%%%%%%%%%
The SSA $A_N$ in single-inclusive hadron production from proton-proton collisions, 
\begin{equation}
p(P, S_{T}) + p(P') \rightarrow h(P_h) + X\,, 
\end{equation}
is defined as
\begin{align}
A_N \equiv \frac{d\Delta\sigma(S_T)}{d\sigma} = 
\frac{\frac{1}{2}\!\left[d\sigma(S_T)-d\sigma(-S_T)\right]}{\frac{1}{2}\!\left[d\sigma(S_T)+d\sigma(-S_T)\right]}.
\label{e:AN}
\end{align}
The numerator and denominator can be expressed, respectively, as~\cite{Qiu:1998ia,Kouvaris:2006zy, Metz:2012ct,Gamberg:2017gle}\footnote{The full asymmetry also involves terms from so-called soft-fermion poles (SFPs) in the tranversely polarized proton~\cite{Koike:2009ge}, a chiral-odd unpolarized twist-3 PDF coupling to transversity~\cite{Kanazawa:2000hz,Zhou:2008mz,Kang:2008ey,Kang:2012em}, and tri-gluon correlators~\cite{Beppu:2013uda}. While the SFP piece might play some role in $A_N$, it cannot account for all of the asymmetry~\cite{Kanazawa:2010au,Kanazawa:2011bg}.  The tri-gluon term has been shown to give small effects in the forward region~\cite{Beppu:2013uda} where $A_N$ is most significant.  The twist-3 unpolarized contribution was shown in Ref.~\cite{Kanazawa:2000kp} to be negligible due to the small size of the corresponding hard partonic cross section.}
\begin{align}
d\Delta \sigma(S_T) &= \frac{2P_{hT}\alpha_S^2}{S}  \displaystyle\sum_i\displaystyle\sum_{a,b,c}\!\displaystyle\int_{z_{min}}^1\!\!\dfrac{dz} {z^3}\displaystyle\int_{x_{min}}^1\!\!\dfrac{dx} {x}\dfrac{1} {x'}\dfrac{1}{xS+U/z} f_1^{b}(x') \bigg[\!M_h\,h_1^{a/p^\uparrow}(x)\,\mathcal{H}^{h/c,i}(x,x',z) \nonumber \\
&\hspace{8.5cm}+\, \dfrac{M}{\hat{u}} \mathcal{F}^{a/p^\uparrow,i}(x,x',z)\,D_1^{h/c}(z)\bigg],\label{e:Num}\\[0.3cm]
d\sigma &= \frac{\alpha_S^2}{S} \displaystyle\sum_i\displaystyle\sum_{a,b,c}\displaystyle\int_{z_{min}}^1\!\dfrac{dz} {z^2}\displaystyle\int_{x_{min}}^1\!\dfrac{dx} {x}\dfrac{1} {x'}\dfrac{1} {xS+U/z}\,f_1^{a/p^\uparrow}(x)\,f_1^{b/p}(x')\,D_1^{h/c}(z)\,S_U^i\,, \label{e:Den}
\end{align}
where $z_{min}=-(T+U)/S$, $x_{min} = -(U/z)/(T/z+S)$, and $x' = -(xT/z)/(xS+U/z)$, with $S = (P+P')^2$, $T = (P-P_h)^2$, and $U = (P'-P_h)^2$.  The summation $\sum_i$ is over all partonic interaction channels; $a$ can be a quark, antiquark, or gluon, and likewise for $b,c$; and $\alpha_s$ is the strong coupling constant.  The hard factors in Eq.~(\ref{e:Den}) for the unpolarized cross section are denoted
by $S_U^i$~\cite{Owens:1986mp,Kang:2013ufa} and can be found in, e.g., Appendix~A of Ref.~\cite{Kouvaris:2006zy}.  
In Eq.~(\ref{e:Num}) the quantities $\mathcal{H}^{h/c,i}(x,x',z)$ and $\mathcal{F}^{a/p^\uparrow,i}(x,x',z)$ are~\cite{Gamberg:2017gle}
\begin{align}
\mathcal{H}^{h/c,i}(x,x',z)& = \left[H_1^{\perp(1),h/c}(z)-z\frac{dH_1^{\perp(1),h/c}(z)} {dz}\right]\tilde{S}_{H_1^{\perp}}^{i}
 +\left[-2H_{1}^{\perp(1),h/c}(z) + \frac{1} {z}\tilde{H}^{h/c}(z)\right] \tilde{S}_{H}^{i}\,, \label{e:mathcalH} \\[0.3cm]
 \mathcal{F}^{a/p^\uparrow,i}(x,x',z) &=  \pi\left[ F_{FT}^{a/p^\uparrow}\!\!(x,x) -x\frac{d F_{FT}^{a/p^\uparrow}\!\!(x,x)} {dx}\right]S^i_{F_{FT}}\,. 
\label{e:mathcalF}
\end{align}
The hard factors $\tilde{S}_{H_1^{\perp}}^i$ and $\tilde{S}_{H}^{i}$ are given in Eqs.~(18)--(23) of Ref.~\cite{Gamberg:2017gle}, while $S_{F_{FT}}^i$ and can be found in Appendix~A of Ref.~\cite{Kouvaris:2006zy}. All hard factors depend on the partonic Mandelstam variables $\hat{s} = xx' S,\,\hat{t} = xT/z,\,{\rm and}\;\hat{u} = x'U/z$.  

There are connections between the non-perturbative functions in Eqs.~(\ref{e:Num})--(\ref{e:mathcalF}) and those in  SIDIS, SIA, and DY.  A generic TMD PDF $F(x,k_T)$ when 
Fourier conjugated into position ($b_T$) space $\tilde{F}(x,b_T)$~\cite{Collins:1981uw,Boer:2011xd,Collins:2011zzd}
exhibits an Operator Product Expansion (OPE) 
in the limit when $b_T$ is small. At leading order in the OPE one has $\tilde{f}_1(x,b_T)\sim f_1(x),\tilde{h}_1(x,b_T)\sim h_1(x),\tilde{f}_{1T}^\perp(x,b_T)\sim \pi F_{FT}(x,x)$, and $\tilde{H}_1^\perp(z,b_T)\sim H_1^{\perp(1)}(z)$~\cite{Kang:2010xv,Aybat:2011ge,Bacchetta:2013pqa,Kanazawa:2015ajw,Gamberg:2017jha,Gutierrez-Reyes:2017glx,Gutierrez-Reyes:2018iod,Scimemi:2018mmi}.
Another way to establish the connection between 
collinear functions and TMDs is by the use of parton model identities\footnote{For discussions on the validity of these relations beyond leading order, see Refs.~\cite{Aybat:2011ge,Kanazawa:2015ajw,Gamberg:2017jha,Scimemi:2019gge,Qiu:2020oqr,Ebert:2022cku}.}:
\begin{align}
f(x)=\int d^2\vec{k}_T f(x,\vec{k}_T^2)\;\;&(f=f_1\,{\rm or}\, h_1)\,,\quad\quad \pi F_{FT}(x,x)=\int \! d^2 \vec{k}_T\,\frac{\vec{k}_T^2} {2M^2} f_{1T}^\perp(x,\vec{k}_T^2)
\equiv f_{1T}^{\perp(1)}(x)\, , \nonumber \\ \label{e:parton_ident}
\\[-0.3cm]
&H_1^{\perp(1)}(z)= z^2\!\int \!d^2 \vec{p}_T \frac{\vec{p}_T^{\,2}} {2M_h^2} H_{1}^\perp(z,z^2\vec{p}_T^{\,2}) \nonumber \,.
\end{align}
We will utilize these relations in parametrizing the TMDs in Sec.~\ref{s:Method}, which allows for a global analysis involving both TMD and CT3 observables.  We also emphasize that the twist-3 FF $\tilde{H}(z)$ in Eq.~(\ref{e:mathcalH}) is the same function that enters Eq.~(\ref{e:sinphiS}).  Using data from HERMES on the $A_{UT}^{\sin\phi_S}$ asymmetry and from BRAHMS and STAR on $A_N^\pi$ then allows us to constrain $\tilde{H}(z)$ for the first time in a global analysis.

%%%%%%%%%%%%%%%%%%%%%%%%%%%%%%%%
%%%%%%%%%%%%%%%%%%%%%%%%%%%%%%%%
\subsection{Nucleon tensor charges and the Soffer bound \label{s:gT_SB}}
%%%%%%%%%%%%%%%%%%%%%%%%%%%%%%%%
%%%%%%%%%%%%%%%%%%%%%%%%%%%%%%%%
The up and down quark tensor charges, $\delta u$ and $\delta d$, of the nucleon, as well as their isovector combination $g_T$, can be computed from the transversity function $h_1(x)$, as
\begin{equation}
\delta u = \int_0^1 dx\,\left(h_1^{u}(x)-h_1^{\bar{u}}(x)\right)\,,\quad\quad \delta d = \int_0^1 dx\,\left(h_1^{d}(x)-h_1^{\bar{d}}(x)\right)\,,\quad\quad g_T\equiv \delta_u - \delta_d\,. \label{e:gT}
\end{equation}
These charges are not only relevant for QCD phenomenology~\cite{Anselmino:2013vqa,Goldstein:2014aja,Radici:2015mwa,Kang:2015msa,Lin:2017stx,Radici:2018iag,Benel:2019mcq,DAlesio:2020vtw,Cammarota:2020qcw} but also for {\it ab initio} studies~\cite{Gupta:2018qil,Yamanaka:2018uud,Hasan:2019noy,Alexandrou:2019brg,Pitschmann:2014jxa} and beyond the Standard Model physics~\cite{Courtoy:2015haa,Yamanaka:2017mef,Gao:2017ade,Gonzalez-Alonso:2018omy}.  The future Electron-Ion Collider will provide crucial measurements to reduce the uncertainties in $\delta u, \delta d$, and $g_T$ so that one can precisely explore these connections~\cite{Gamberg:2021lgx}.  Lattice QCD computations at the physical point of the tensor charges in Eq.~(\ref{e:gT})~\cite{Gupta:2018qil,Yamanaka:2018uud,Hasan:2019noy,Alexandrou:2019brg} provide important constraints on $h_1(x)$ in addition to the experimental observables discussed above.  In fact, lattice calculations of $h_1$ itself as a function of $x$ even exist using the approach of  pseudo-PDFs~\cite{HadStruc:2021qdf} or quasi-PDFs~\cite{Alexandrou:2019lfo,Alexandrou:2021oih}. We also mention that the tensor charges are almost entirely a valence effect.  The lattice calculations in Refs.~\cite{Gupta:2018qil,Alexandrou:2019brg} show that disconnected diagrams give contributions that are about two orders of magnitude smaller than those from connected diagrams.  In addition, if one assumes a symmetric sea, then $g_T$ in Eq.~(\ref{e:gT}) only involves the up and down valence transversity PDFs.

The probability interpretation of $h_1(x)$ also allows for a theoretical constraint to be derived, the so-called Soffer bound~\cite{Soffer:1994ww}:
\begin{equation}
    \left|h_1^q(x)\right|\le \frac{1}{2}\left(f_1^q(x)+g_1^q(x)\right), \label{e:SB}
\end{equation}
where $g_1(x)$ is the helicity PDF.  The Soffer bound has been explicitly imposed in many previous extractions of transversity~\cite{Anselmino:2013vqa,Radici:2015mwa,Kang:2015msa,Radici:2018iag,DAlesio:2020vtw} but not in {\tt JAM3D-20}~\cite{Cammarota:2020qcw}. Note that the validity of the Soffer bound, and the consequences of its violation, were discussed in the past in the context of $Q^2$ dependence in Ref.~\cite{Goldstein:1995ek}  and even of confinement in Ref.~\cite{Ralston:2008sm}. In addition, recent work~\cite{Collins:2021vke} suggests that positivity bounds do not hold beyond a parton model picture. Nevertheless, since the experimental indications of any violation are unclear, it is useful at this stage to study the impact of the Soffer bound on our global analysis of SSAs.

%%%%%%%%%%%%%%%%%%%%%%%%%%%%%%%%%%%%%%%%%%%%%%%%%%%%%%%%%%%%%%%%%%%%
\section{Phenomenological Results}\label{s:Pheno}
%%%%%%%%%%%%%%%%%%%%%%%%%%%%%%%%%%%%%%%%%%%%%%%%%%%%%%%%%%%%%%%%%%%%
Our update to the global analysis of Ref.~\cite{Cammarota:2020qcw} further explores and extends upon several aspects of those results. Prior to {\tt JAM3D-20}, most phenomenological extractions of the tensor charges based on Eq.~(\ref{e:gT}) found tension with lattice QCD calculations~\cite{Anselmino:2013vqa,Radici:2015mwa,Kang:2015msa,Radici:2018iag,Benel:2019mcq,DAlesio:2020vtw}.  The work in Ref.~\cite{Lin:2017stx} demonstrated that including lattice data on $g_T$ along with Collins effect SIDIS data alleviated that discrepancy. In {\tt JAM3D-20} we found for the first time (without including lattice data) agreement with the lattice QCD values for $\delta u, \delta d$, and $g_T$.  The $A_N^\pi$ data from proton-proton collisions was the key driver in pushing the tensor charges, especially $\delta u$, to the lattice values, whereas typically phenomenology undershoots $\delta u$ and $g_T$~\cite{Anselmino:2013vqa,Radici:2015mwa,Kang:2015msa,Radici:2018iag,Benel:2019mcq,DAlesio:2020vtw}\footnote{We recall as well that the term involving $\tilde{H}(z)$ in the fragmentation piece of $A_N^\pi$ was ignored.}.  This was one of the main outcomes of {\tt JAM3D-20} and highlights the importance of including all SSA measurements in a global analysis.  

At the same time, the extracted $h_1(x)$ from {\tt JAM3D-20} showed a clear violation of the Soffer bound, especially for the down quark.  The question then arises as to whether or not the Soffer bound is compatible with $A_N$ data and lattice QCD tensor charge computations.  
Therefore, in this update we include the same experimental data as {\tt JAM3D-20} (Sivers and Collins asymmetries in SIDIS from HERMES and COMPASS~\cite{ Alekseev:2008aa,Adolph:2014zba,HERMES:2020ifk}, Collins effect in SIA from Belle, BaBar, and BESIII~\cite{Seidl:2008xc,TheBABAR:2013yha,Aubert:2015hha,Ablikim:2015pta}, Sivers effect in DY from STAR and COMPASS~\cite{Adamczyk:2015gyk,Aghasyan:2017jop}, and $A_N^\pi$ from BRAHMS and STAR~\cite{Lee:2007zzh,Adams:2003fx,Abelev:2008af,Adamczyk:2012xd}), except for the following changes:~(i) HERMES data~\cite{Airapetian:2009ae,Airapetian:2010ds} has been replaced with their superseding 3D-binned measurements~\cite{HERMES:2020ifk}; (ii) data on $A_{UT}^{\sin\phi_S}$~\cite{HERMES:2020ifk} is included to allow for an extraction of $\tilde{H}(z)$, which was previously set to zero in {\tt JAM3D-20}; (iii) the Soffer bound on transversity is imposed; (iv) the lattice tensor charge $g_T$ value from Ref.~\cite{Alexandrou:2019brg} is now included in the analysis as a data point using Eq.~(\ref{e:gT}).

%%%%%%%%%%%%%%%%%%%%%%%%%%%%%%%%
%%%%%%%%%%%%%%%%%%%%%%%%%%%%%%%%
\subsection{Methodology}\label{s:Method}
%%%%%%%%%%%%%%%%%%%%%%%%%%%%%%%%
%%%%%%%%%%%%%%%%%%%%%%%%%%%%%%%%
We review here the procedure used in {\tt JAM3D-20} to extract the relevant non-perturbative functions from a global analysis. We maintain the same features of that study in order to clearly identify changes associated with additional constraints from the Soffer bound on transversity, lattice QCD tensor charge data, and the $A_{UT}^{\sin\phi_S}$ measurement from HERMES.
Therefore, we employ a Gaussian ansatz in transverse momentum space and decouple the $x$ and $k_T$ ($z$ and $p_T$) dependence. Although this type of parametrization does not have the complete features of
TMD evolution, it was shown in 
Refs.~\cite{Anselmino:2016uie,Anselmino:2015sxa}
that utilizing such a parametrization is comparable to full TMD evolution at next-to-leading-logarithmic accuracy~\cite{Sun:2013dya,Kang:2014zza,Kang:2015msa,Echevarria:2014xaa,Kang:2017btw}.  
In addition, asymmetries are ratios of cross sections 
where evolution and next-to-leading order effects tend to cancel out~\cite{Kang:2017btw}. 
For the unpolarized and transversity TMDs we have\footnote{Any PDF where the nucleon is not explicitly indicated is assumed to be for a proton.  Also, $q$ can represent either a quark or antiquark.}
\begin{eqnarray}
f^q(x,\vec{k}_T^2) 
&=& f^q(x)\ {\cal G}_{f}^q(k_T^2)\,,
\label{eq:f1h1-gauss}
\end{eqnarray}
where the generic function $f = f_1$ or $h_1$, and
\begin{eqnarray}
{\cal G}_f^q(k_T^2)
&=& \frac{1}{\pi\avkT_f^q}\;
{\exp\!\left[{-\frac{\kT^2}{\avkT_f^q}}\right]},
\label{eq:Ggauss}
\end{eqnarray}
with $k_T\equiv |\vec{k}_T|$. Using the relation  
$\pi F_{FT}(x,x)=f_{1T}^{\perp(1)}(x)$~\cite{Boer:2003cm} from Eq.~(\ref{e:parton_ident}), 
the Sivers function reads
\begin{eqnarray}
f_{1T}^{\perp \,q}(x,\vec{k}_T^2)
&=& \frac{2 M^2}{\avkT^{q}_{f_{1T}^\perp}}\,
    \pi F_{FT}(x,x)\
    {\cal G}_{f_{1T}^\perp}^{q}\!(k_T^2)\,.
\label{e:sivers}
\end{eqnarray}
For the TMD FFs, the unpolarized function is parametrized as
\begin{eqnarray}
D_1^{h/q}(z,z^2\vec{p}_T^{\,2})
&=& D_1^{h/q}(z)\ {\cal G}_{D_1}^{h/q}(z^2p_T^2)\,,
\end{eqnarray}
while the Collins FF reads
\begin{eqnarray}
H_1^{\perp h/q}(z,z^2\vec{p}_T^{\,2})
&=& \frac{2 z^2 M_h^2}{\avpperp^{h/q}_{H_1^\perp}}\,
    H_{1\, h/q}^{\perp (1)}(z)\
    {\cal G}_{H_1^\perp}^{h/q}(z^2p_T^2)\,,
\label{e:collins}
\end{eqnarray}
where we have explicitly written its $z$ dependence in terms of its
first moment $H_{1\, h/q}^{\perp (1)}(z)$~\cite{Kang:2015msa}. The widths for the FFs are denoted as $\avpperp^{h/q}_{D}$, where $D=D_1\,{\rm or}\, H_1^\perp$. (Note that the hadron transverse momentum $\vec{P}_\perp$ with respect to the fragmenting quark is $\vec{P}_\perp = -z\vec{p}_T$.) 
For $f_1^q(x)$ and $D_1^{h/q}(z)$ we use the leading-order
CJ15~\cite{Accardi:2016qay} and DSS~\cite{deFlorian:2007ekg}
functions, following the same set up as in {\tt JAM3D-20}.  The pion PDFs  are taken from Ref.~\cite{Barry:2018ort} and are next-to-leading order.
We generically parametrize the collinear functions  $h_1(x)$, $F_{FT}(x,x)$, $H_1^{\perp(1)}(z)$, $\tilde{H}(z)$, at an initial scale of $Q_0^2=2\,{\rm GeV^2}$, as
\begin{equation}
F^q(x)\!=\!\frac{N_q\,x^{a_q}(1-x)^{b_q}(1+\gamma_q\,x^{\alpha_q}(1-x)^{\beta_q})}
            {{\rm B}[a_q\!+\!2,b_q\!+\!1]
             +\gamma_q {\rm B}[a_q\!
             +\!\alpha_q\!+\!2,b_q\!
             +\!\beta_q\!+\!1]}   
\,, 
\label{e:colfunc}
\end{equation}
where $F^q=h_1^q,\pi F_{FT}^q$, $H_{1 \,h/q}^{\perp (1)}$, $\tilde{H}^{h/q}$ (with $x\to
z$ for the latter two functions), and B is the Euler beta function.  We also implement a DGLAP-type evolution for the collinear part of these functions,  
analogous to Ref.~\cite{Duke:1983gd}, where a
double-logarithmic $Q^2$-dependent term is explicitly added to the
parameters, just as was done in {\tt JAM3D-20}.  For the unpolarized collinear twist-2 PDFs and FFs, we use the standard leading-order DGLAP evolution.  The scale at which the non-perturbative functions are evaluated in calculating an observable is set by the hard scale $Q^2$ of the particular process.  SIDIS data has $2 \lesssim Q^2\lesssim 40\,{\rm GeV^2}$, while  
SIA data has $Q^2\approx 13\,{\rm GeV^2}$ for BESIII 
or $110\, {\rm GeV^2}$ for Belle and BaBar. 
For DY data, $Q^2\approx 30\,{\rm GeV^2}$ for COMPASS or $\approx(80\,{\rm GeV})^2$ for STAR, while $A_N^{\pi}$ data has $1\lesssim Q^2\lesssim 6\,{\rm GeV^2}$ for BRAHMS and  $1\lesssim Q^2\lesssim 13\,{\rm GeV^2}$ for STAR.

For the collinear PDFs $h_1^q(x)$ and $\pi F_{FT}^q(x,x)$, we only
allow $q=u,d$ and set antiquark functions to zero. Nevertheless, the $u$ and $d$ functions are understood as being the sum of valence and sea contributions, i.e., $u=u_v+\bar{u}$ and $d=d_v+\bar{d}$.
For both functions, $\{\gamma,\alpha,\beta\}$ are not used, and we  set $b_u=b_d$, as the $\chi^2/{\rm npts}$ does not improve by leaving more parameters free.  
For the collinear FFs $H_{1 \,h/q}^{\perp (1)}(z)$ and $\tilde{H}^{h/q}(z)$, we allow for
favored ($fav$)  and unfavored ($unf$) parameters, with $fav$ corresponding to the fragmentation channels $u\to\pi^+$, $\bar d\to\pi^+$ ($\bar u\to\pi^-$, $d\to\pi^-$) and  $unf$ for all other flavors. 
For $H_{1 \,h/q}^{\perp (1)}(z)$, exactly as in {\tt JAM3D-20}, $\{\gamma, \beta\}$ are free while $\alpha$ is set to zero.  This is due to the change in shape of the SIA data as a function of $z$ and the fact that the data are at larger $z>0.2$. For $\tilde{H}^{h/q}(z)$, $\{\gamma,\alpha,\beta\}$ are not used, and we set $a_{fav}=a_{unf}$ and $b_{fav}=b_{unf}$.  We  have verified that no meaningful change in the $\chi^2/{\rm npts}$ occurs if $a$ and $b$ are separately fit for favored and unfavored, as the SIDIS and $A_N$ data are not sensitive enough to $\tilde{H}^{h/q}(z)$ to constrain more free parameters.
In the end we have a total of 24 parameters for the collinear functions.  
There are also 4 parameters for the transverse momentum widths associated with $h_1$, $f_{1T}^\perp$, and $H_{1}^{\perp}$: $\avkT^{u}_{f_{1T}^\perp}=\avkT^{d}_{f_{1T}^\perp}\equiv\avkT_{f_{1T}^\perp}$; 
$\avkT_{h_1}^u=\avkT_{h_1}^d\equiv \avkT_{h_1}$;     $\avpperp^{fav}_{H_1^\perp}$ and
$\avpperp^{unf}_{H_1^\perp}$.
We extract the unpolarized TMD widths~\cite{Anselmino:2005nn,Signori:2013mda,Anselmino:2013lza} by including HERMES
pion and kaon multiplicities~\cite{Airapetian:2012ki}, which involves 6 more parameters:~$\avkT^{val}_{f_1}$,
$\avkT^{sea}_{f_1}$, $\avpperp^{fav}_{D_1^{\pi}}$,
$\avpperp^{unf}_{D_1^{\pi}}$,$\avpperp^{fav}_{D_1^{K}}$,
$\avpperp^{unf}_{D_1^{K}}$. Our working hypothesis for the pion PDF widths, like in {\tt JAM3D-20}, is that they are the same as those for the proton.  This is roughly supported by a Dyson-Schwinger calculation evolved to typical energy scales of DY~\cite{Shi:2018zqd}, but more detailed analyses of pion TMDs may yield different results~\cite{Vladimirov:2019bfa}, and further studies are needed.  Since only the Sivers COMPASS DY data, which have rather large uncertainties, involve pion TMDs, our assumption should be sufficient for this analysis. 
The JAM Monte Carlo framework~\cite{Cammarota:2020qcw} is used to sample the Bayesian posterior distribution with approximately 500  replicas of the parameters in order to estimate uncertainties for our extracted non-perturbative quantities.  

We now comment on how we enforce the Soffer bound (SB) in Eq.~(\ref{e:SB}). Several prior analyses have explicitly used the r.h.s.~of~Eq.~(\ref{e:SB}) in the parametrization of $h_1(x)$~\cite{Anselmino:2013vqa,Radici:2015mwa,Kang:2015msa,Radici:2018iag,DAlesio:2020vtw}.  However, since the parametrization of the collinear functions in our analysis generically follows Eq.~(\ref{e:colfunc}), such a route is not feasible.  Therefore, we instead generate ``data'' at a scale of $Q_0^2 = 2\,{\rm GeV^2}$ for the r.h.s.~of Eq.~(\ref{e:SB}) for $0< x<1$ using the unpolarized and helicity PDFs from Ref.~\cite{Cocuzza:2022jye}.\footnote{If the SB holds at some initial scale, then evolution will not cause a violation~\cite{Kamal:1996gqi,Barone:1997fh,Vogelsang:1997ak}.  This justifies only using data at $Q_0^2$.}  The fact that $f_1(x)$ and $g_1(x)$ were extracted in Ref.~\cite{Cocuzza:2022jye} simultaneously using Monte Carlo methods allows us to use their replicas to calculate a central value and 1-$\sigma$ uncertainty for the r.h.s.~of (\ref{e:SB}) at a given $x$.  This SB data is then included in our analysis as an additional constraint. However, the theory calculation of $|h_1(x)|$, point-by-point in $x$,  only contributes to the overall $\chi^2$ if the l.h.s.~of Eq.~(\ref{e:SB}) violates the inequality with the generated SB data on the r.h.s.~by more than 1-$\sigma$. Other groups have also explored relaxing a strict imposition of the Soffer bound~\cite{Benel:2019mcq,DAlesio:2020vtw}. 

Lastly, since HERMES released 3D-binned measurements that supersede their previous data, we performed the {\tt JAM3D-20} analysis again but now using the new HERMES data for the Sivers and Collins effects.  This update to {\tt JAM3D-20} will be referred to as {\tt JAM3D-20+} . The comparison of the non-perturbative functions from {\tt JAM3D-20} and {\tt JAM3D-20+}  is shown in Fig.~\ref{f:qcf_JAM20plus} of Appendix~\ref{s:app_a}. The only noticeable change is the Sivers function has become larger and falls off more slowly at larger $x$, although the relative uncertainty remains the same.

%%%%%%%%%%%%%%%%%%%%%%%%%%%%%%%%
%%%%%%%%%%%%%%%%%%%%%%%%%%%%%%%%
\subsection{Extraction of Non-Perturbative Functions and Tensor Charges}\label{s:Results}
%%%%%%%%%%%%%%%%%%%%%%%%%%%%%%%%
%%%%%%%%%%%%%%%%%%%%%%%%%%%%%%%%
\begin{table*}[t]
\centering
\begin{tabular}{ |c|c|c|c|c| }
 \hline
{\bf Observable} & 
{\bf Reactions} & 
{\bf Non-Perturbative Function(s)} &  
$\boldsymbol{\chi^2}/\boldsymbol{{\rm npts}}$ &
{\bf Exp.~Refs.\!} 
\\\hline \hline

$A_{UT}^{\sin(\phi_h-\phi_S)}$  &
$e+(p,d)^\uparrow\to e+(\pi^+,\pi^-,\pi^0)+X$  & 
$f^{\perp}_{1T}(x,\vec{k}_T^2)$  & 
$\! 182.9/166=1.10 \!$ & 
\cite{HERMES:2020ifk, Alekseev:2008aa,Adolph:2014zba}  
\\

$A_{UT}^{\sin(\phi_h+\phi_S)}$  &
$e+(p,d)^\uparrow\to e+(\pi^+,\pi^-,\pi^0)+X$  & 
$h_{1}(x,\vec{k}_T^2),H_1^{\perp}(z,z^2 \vec{p}_T^{\,2})$  & 
$\! 181.0/166=1.09 \!$ & 
\cite{HERMES:2020ifk, Alekseev:2008aa,Adolph:2014zba} 
\\

$^{*\!\!}A_{UT}^{\sin\phi_S}$  &
$e+p^\uparrow\to e+(\pi^+,\pi^-,\pi^0)+X$  & 
$h_{1}(x),\tilde{H}(z)$  & 
$18.6/36=0.52$ & 
\cite{HERMES:2020ifk, Alekseev:2008aa,Adolph:2014zba} 
\\\hline

$A_{UC/UL}$  &
$e^++e^-\to \pi^+\pi^-(UC,UL)+X$  & 
$H_1^{\perp}(z,z^2 \vec{p}_T^{\,2})$  & 
$154.9/176=0.88$ & 
\cite{Seidl:2008xc,TheBABAR:2013yha,Aubert:2015hha,Ablikim:2015pta} 
\\\hline

$A_{T,\mu^+\mu^-}^{\sin\phi_S}$  &
$\pi^-\!+p^\uparrow\to \mu^+\mu^-+X$  & 
$f_{1T}^{\perp}(x,\vec{k}_T^2)$  & 
$6.92/12=0.58$ & 
\cite{Aghasyan:2017jop} 
\\[0.1cm]

$A_N^{W/Z}$  &
$p^\uparrow +p \to (W^+,W^-,Z)+X$  & 
$f_{1T}^{\perp}(x,\vec{k}_T^2)$  & 
$30.8/17=1.81$ & 
\cite{Adamczyk:2015gyk} 
\\
 
$A_N^\pi$  &
$p^\uparrow + p\to (\pi^+,\pi^-,\pi^0) + X$  & 
$\! h_1(x),F_{FT}(x,x)= \tfrac{1} {\pi}f_{1T}^{\perp(1)}(x),H_1^{\perp(1)}(z),\tilde{H}(z)\!$  & 
$70.4/60=1.17$ & 
\cite{Lee:2007zzh,Adams:2003fx,Abelev:2008af,Adamczyk:2012xd} 
\\ \hline 

${\rm Lattice\;}g_T$  &
------  & 
$h_1(x)$  & 
$1.82/1=1.82$ & \cite{Alexandrou:2019brg}
\\ \hline
\end{tabular}
\caption{
    Summary of the observables analyzed in {\tt JAM3D-22} .  There are a
    total of 21 different reactions.
    There are also a total of 8 non-perturbative functions when one takes into account flavor separation. The $\chi^2$ is computed based on calculating for each point the theory expectation value from the replicas. $^*$For the $A_{UT}^{\sin\phi_S}$ data we only use the $x$- and $z$-projections. \vspace{-0.5cm}} 
\label{t:sum}
\end{table*}
\subsubsection{{\tt JAM3D-22}  non-perturbative functions}
\begin{figure}[t!]
\includegraphics[width=0.7\textwidth]{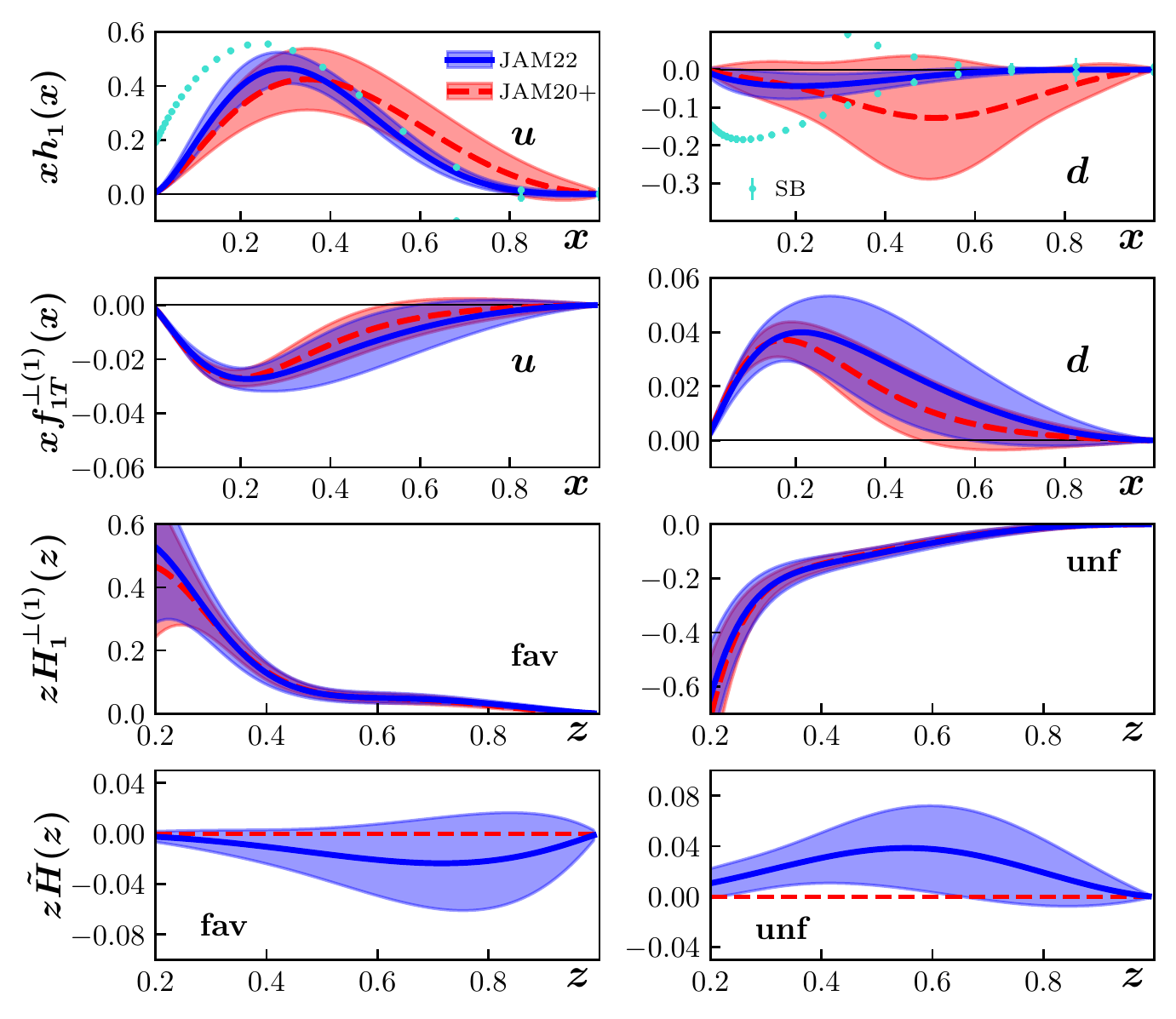}\vspace{-0.2cm}
\caption{The extracted functions $h_1(x)$, $f_{1T}^{\perp(1)}(x)$, $H_1^{\perp(1)}(z)$, and $\tilde{H}(z)$ at $Q^2=4$ GeV$^2$ from our {\tt JAM3D-22}  global analysis (blue solid curves with 1-$\sigma$ CL error bands) compared to {\tt JAM3D-20+}  global
analysis (red dashed curves with 1-$\sigma$ CL error bands). The generated Soffer bound (SB) data are also displayed (cyan points). \vspace{-0.25cm}} 
\label{f:qcf}
\end{figure}

Our update to the simultaneous global analysis of SSAs ({\tt JAM3D-22}) includes all the observables in Table~\ref{t:sum}.  The cuts from {\tt JAM3D-20} of $0.2 < z< 0.6,\; Q^2>1.63\,{\rm GeV^2}, \;{\rm and}\;0.2 <P_{hT}<0.9 \,{\rm GeV}$ 
have been applied to all SIDIS data sets and $P_{hT}>1\,{\rm GeV}$ to all $A_N^\pi$ data sets.  At this stage we do not explore the effect of what different cuts or criteria~\cite{Bacchetta:2017gcc,Scimemi:2019cmh,Echevarria:2020hpy,Bury:2021sue,Boglione:2022gpv} for the applicability of TMD factorization have on our fit.  The non-perturbative functions extracted from our analysis\footnote{A {\tt Google Colab} notebook that allows one to generate the functions and asymmetries from our analysis can be found in Ref.~\cite{jam3dlib}.} are shown in Fig.~\ref{f:qcf} compared to {\tt JAM3D-20+}.  The most noticeable change is in the shape of $h_1(x)$, which is due to the fact that we now include Soffer bound data in our fit (as described towards the end of Sec.~\ref{s:Method}).  The up quark transversity function is not as large as {\tt JAM3D-20+} and falls off slightly faster at larger $x$.  The down quark transversity experiences the most significant changes, peaking at lower $x$ with about half the magnitude from before and falling off much quicker at larger $x$.  

We also show in Fig.~\ref{f:qcf} our extraction of $\tilde{H}(z)$.  The sign is in agreement with the fit of Ref.~\cite{Kanazawa:2014dca} and model calculation of Ref.~\cite{Lu:2015wja}, and the magnitude is also reasonable given these previous studies.  Similar to the Collins FF, favored and unfavored $\tilde{H}(z)$ have a preference to be opposite in sign and roughly equal in magnitude.  This is not unexpected given that both $H_1^{\perp(1)}(z)$ and $\tilde{H}(z)$ are derived from the same underlying quark-gluon-quark correlator~\cite{Kanazawa:2015ajw}. 

We further explored the significance of the signal obtained for $\tilde{H}^{fav}(z)$ and $\tilde{H}^{unf}(z)$ by setting one of them to zero and re-running the analysis.  The case where $\tilde{H}^{fav}(z)=0$ showed almost no change in the $\chi^2/{\rm npts}$ for $A_{UT}^{\sin\phi_S}$, while the $\chi^2/{\rm npts}$ for $A_N^\pi$ increases slightly  from 1.17 to 1.27.  When $\tilde{H}^{unf}(z)=0$, the $\chi^2/{\rm npts}$ for $A_{UT}^{\sin\phi_S}$ increases to 1.58 while the $\chi^2/{\rm npts}$ for $A_N^\pi$ remains basically the same. This shows that $A_{UT}^{\sin\phi_S}$ is mainly driven by the unfavored fragmentation channels $d\to \pi^+$ and $u\to \pi^-$, especially the $\pi^-$ final state that has a clear nonzero signal (see Fig.~\ref{f:sidis_sinphiS}), whereas the favored fragmentation channels $u\to \pi^+$ and $d\to \pi^-$ from the $\tilde{H}$ term in $A_N^\pi$ help in better describing those data.

Another significant feature of {\tt JAM3D-22} is the reduction in the error band for $h_1(x)$.  This can be partly attributed to imposing the Soffer bound but even more so due to the inclusion of the lattice $g_T$ data point. In Fig.~2 we provide a comparison of $h_1(x)$ between our full {\tt JAM3D-22}  analysis (which includes both lattice data and the Soffer bound) and one with no lattice data ({\tt JAM3D-22}  no LQCD).  One can see that including the lattice $g_T$ point in the fit causes about a $50\%$ reduction in the uncertainty for $h_1(x)$. This further highlights the impact of the Soffer bound and lattice $g_T$ in constraining the function. There is also an effect on $\tilde{H}(z)$, which is due to the fact that $h_1(x)$ couples to $\tilde{H}(z)$ in the $A_{UT}^{\sin\phi_S}$ and $A_N^\pi$ asymmetries (see Eqs.~(\ref{e:sinphiS}) and (\ref{e:Num}), (\ref{e:mathcalH})).  Therefore, placing tighter constraints on transversity by including lattice data allows for a slightly more precise extraction of $\tilde{H}(z)$.  

We explored using a more flexible parametrization for $h_1(x)$ (see also Refs.~\cite{Bacchetta:2012ty,Benel:2019mcq}), where $\gamma_u$, $\gamma_d$, $\alpha_u=\alpha_d$, and $\beta_u=\beta_d$ in Eq.~(\ref{e:colfunc}) were free parameters, but did not find any change to the transversity error band or $\chi^2/{\rm npts}$ from the aforementioned analyses. In fact, the parametrization (\ref{e:colfunc}) without $\gamma,\alpha,\beta$ is already very flexible due to the nature of the Monte Carlo sampling procedure used in our analysis, which generates a large family of functions as priors.  The topic of properly quantifying PDF/FF uncertainties has been given its own specialized attention in the literature, and we point the reader to dedicated studies in Refs.~\cite{Watt:2012tq,Rojo:2015acz,Hou:2016sho,Carrazza:2016htc,Ball:2022hsh,Courtoy:2022ocu,Hunt-Smith:2022ugn} and references therein.

We mention that $f_{1T}^{\perp(1)}(x)$ and $H_1^{\perp(1)}(z)$ are essentially identical between the two fits ({\tt JAM3D-22}  and {\tt JAM3D-22}  no LQCD).  This demonstrates that, although the Sivers function can be influenced by transversity due the fact that both enter $A_N^\pi$, the main constraint on $f_{1T}^{\perp(1)}(x)$ is from the Sivers effects in SIDIS and DY.  Likewise, even though $h_1(x)$ couples to $H_1^{\perp(1)}(z)$ in the Collins effect in SIDIS and $A_N^\pi$ fragmentation term, the Collins effect in SIA has the most significant impact on the Collins function's size and shape.

\begin{figure}[h]
\includegraphics[width=0.7\textwidth]{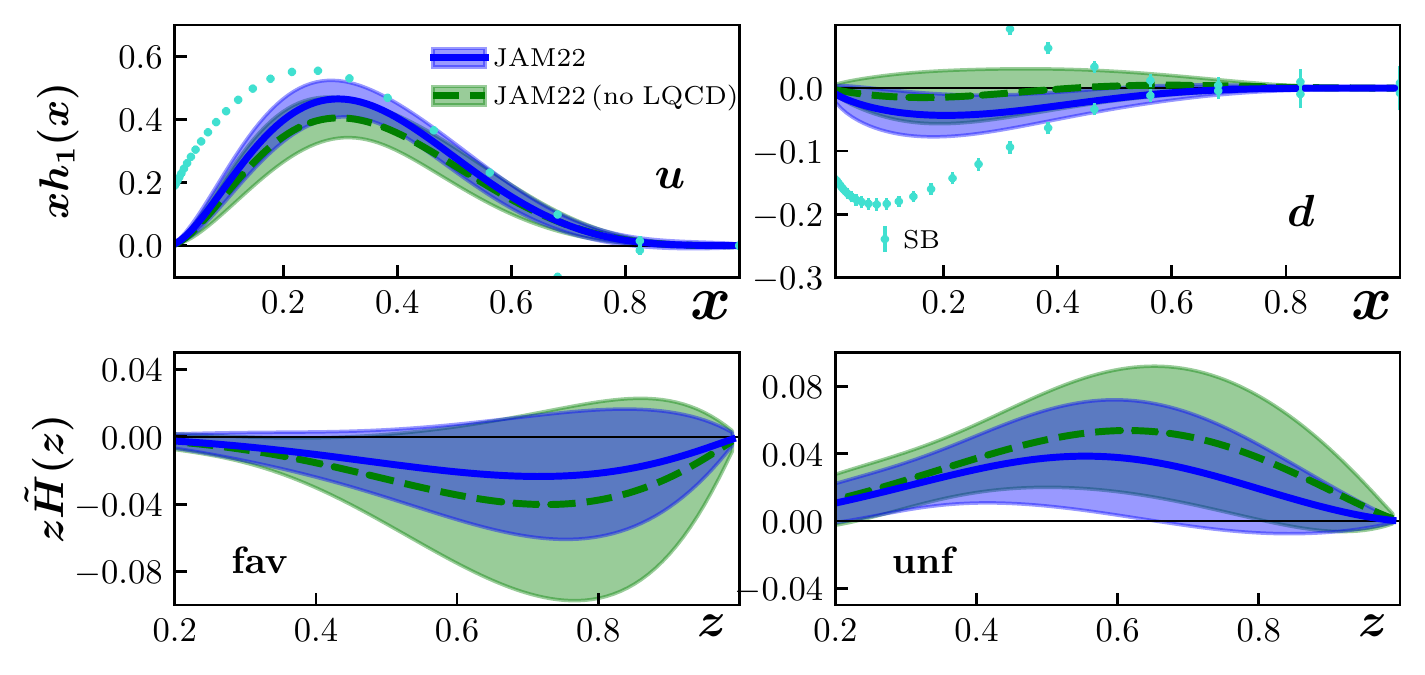}\vspace{-0.2cm}
\caption{The extracted functions $h_1(x)$ and $\tilde{H}(z)$ at $Q^2=4$ GeV$^2$ from our {\tt JAM3D-22}  global analysis (blue solid curves with 1-$\sigma$ CL error bands) compared to a fit without lattice QCD data (green dashed curves with 1-$\sigma$ CL error bands). The generated Soffer bound data are also displayed (cyan points). The functions $f_{1T}^{\perp(1)}(x)$ and $H_1^{\perp(1)}(z)$ are essentially identical between the two fits, so we do not show them here.\vspace{-0.25cm}} 
\label{f:qcf_SB_LatgT}
\end{figure}

\subsubsection{Tensor charges}
\begin{figure}[t]
\centering
\includegraphics[width=0.7\textwidth]{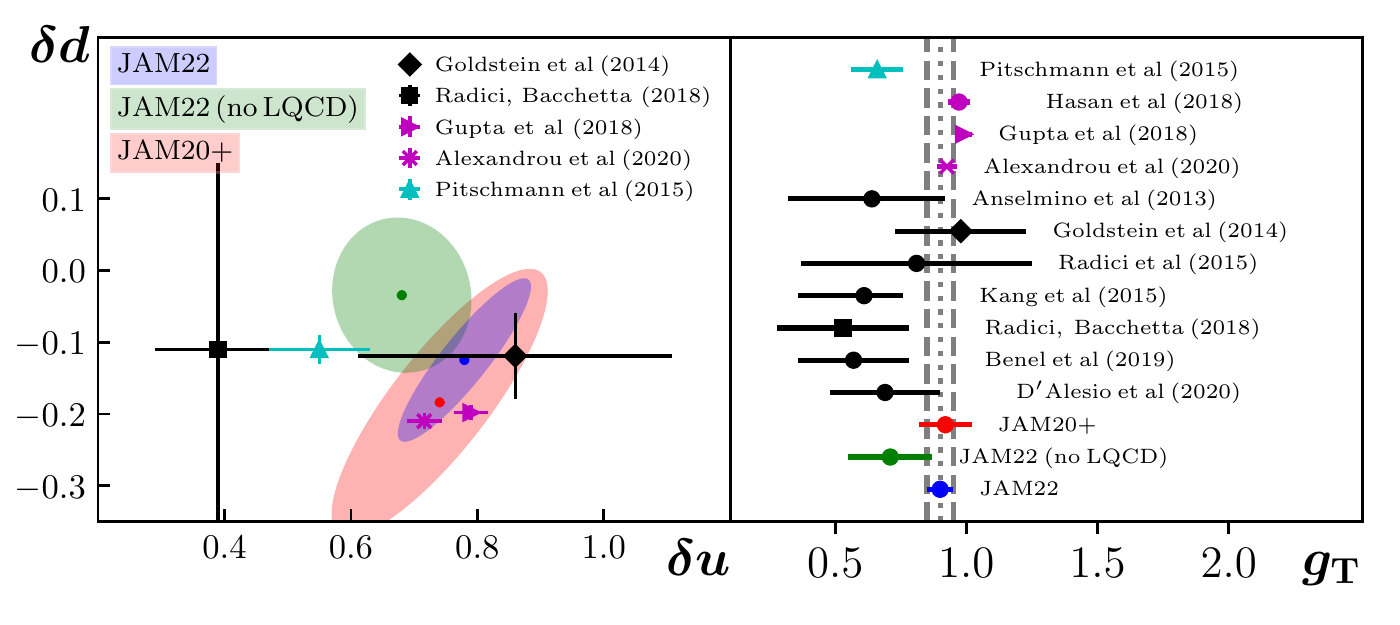}\vspace{-0.5cm}
\caption{
The tensor charges $\delta u$, $\delta d$, and $g_T$. Our new {\tt JAM3D-22} results (blue) are compared to an analysis that does not include the lattice $g_T$ data point ({\tt JAM3D-22} no LQCD in green) and to the {\tt JAM3D-20+} results (red)  at
$Q^2=4$~GeV$^2$ along with others from phenomenology (black), lattice QCD (purple), and
Dyson-Schwinger (cyan).
\vspace{-0.3cm}} 
\label{f:gT}
\end{figure}

From the transversity function we are able to calculate the tensor charges $\delta u, \delta d$, and $g_T$ using Eq.~(\ref{e:gT}).  For {\tt JAM3D-22}  we find $\delta u=0.78\pm 0.11, \delta d=-0.12\pm 0.11$, and $g_T=0.90\pm 0.05$. These results are shown in Fig.~\ref{f:gT} compared to an analysis that does not include the lattice $g_T$ data point ({\tt JAM3D-22}  no LQCD) and the {\tt JAM3D-20+} fit as well as the computations from other phenomenological and lattice studies. The inclusion of the precise lattice QCD data point for $g_T$ from Ref.~\cite{Alexandrou:2019brg} (Alexandrou, et al.~(2020) in Fig.~\ref{f:gT}) causes a substantial reduction in the uncertainty for $\delta u, \delta d$, and $g_T$ between {\tt JAM3D-20+} and {\tt JAM3D-22}.  

{\tt JAM3D-20/20+}~\cite{Cammarota:2020qcw} was the first phenomenological analysis to find agreement with lattice values for the tensor charges.  Previous extractions~\cite{Anselmino:2013vqa,Radici:2015mwa,Kang:2015msa,Radici:2018iag,Benel:2019mcq,DAlesio:2020vtw} typically fell below the lattice results for $\delta u$ and $g_T$, even when relaxing the Soffer bound constraint~\cite{Benel:2019mcq,DAlesio:2020vtw}.  In {\tt JAM3D-20/20+}, the $A_N^\pi$ data drove the magnitude of $h_1^u(x)$ and $h_1^d(x)$ to be larger than studies that only included TMD or dihadron observables, highlighting the importance of a global analysis of SSAs.  The imposition of the Soffer bound in our {\tt JAM3D-22}  global analysis restricts the size of transversity, especially for the down quark.  In addition, now that the $\tilde{H}(z)$ term in $A_N^\pi$ (see Eq.~(\ref{e:mathcalH})) is not set to zero, $h_1^u(x)$ and $h_1^d(x)$ do not need to be as large in order to achieve agreement with the $A_N^\pi$ data.  Consequently, if one does {\it not} include lattice data in the analysis ({\tt JAM3D-22}  no LQCD), the values for $\delta u, \delta d$, and $g_T$ become smaller than in the {\tt JAM3D-20/20+} case, as one sees in Fig.~\ref{f:gT}.  The value for $\delta u$ still agrees with lattice within uncertainties, but $\delta d$ and $g_T$ are about 1- to 1.5-$\sigma$ below.  

However, when the lattice $g_T$ data point is included, as in the full {\tt JAM3D-22}  scenario, then one again finds agreement with the lattice results, with $h_1^u(x)$ and $h_1^d(x)$ increasing in magnitude accordingly -- see Fig.~\ref{f:qcf_SB_LatgT}.  This fact conveys an important point:~an analysis, at a superficial glance, may appear to have tension with the lattice tensor charge values, but one cannot definitively determine this until lattice data is included.  A similar conclusion was found in Ref.~\cite{Lin:2017stx}.  That is, the  analysis may be able to find solutions that are compatible with both lattice and experimental data maintaining an acceptable value for the $\chi^2/{\rm npts}$. 

We also mention that the behavior/uncertainty of the transversity PDF below $x\sim 0.01$ (the lowest $x$ for which there is data) does not affect the previous conclusions.  Specifically, we calculated the truncated moments of the tensor charges, integrating only down to $x=0.01$, and found that $\delta u=0.78\pm 0.10, \delta d=-0.11\pm 0.09, g_T=0.89\pm 0.05$, which are almost exactly the values of the full moments.
Actually, small $x$ (below $x\sim 0.01$) is a region where different dynamics set in, and one should resum logs of $x$ rather than $Q^2$. These small-$x$ evolution equations remove the extrapolation bias inherent in parametrizing the $x$ dependence because they are able to {\it predict} the small-$x$ behavior from first principles.  Such a study was carried out recently for the helicity PDF~\cite{Adamiak:2021ppq}.  Analogously, if one wants to rigorously handle the small-$x$ region for transversity, then the relevant small-$x$ evolution equations derived in Ref.~\cite{Kovchegov:2018zeq} should be implemented into the analysis, which is beyond the scope of this work.

\subsubsection{Comparison with other groups}

\begin{figure}[h!]
\includegraphics[width=0.7\textwidth]{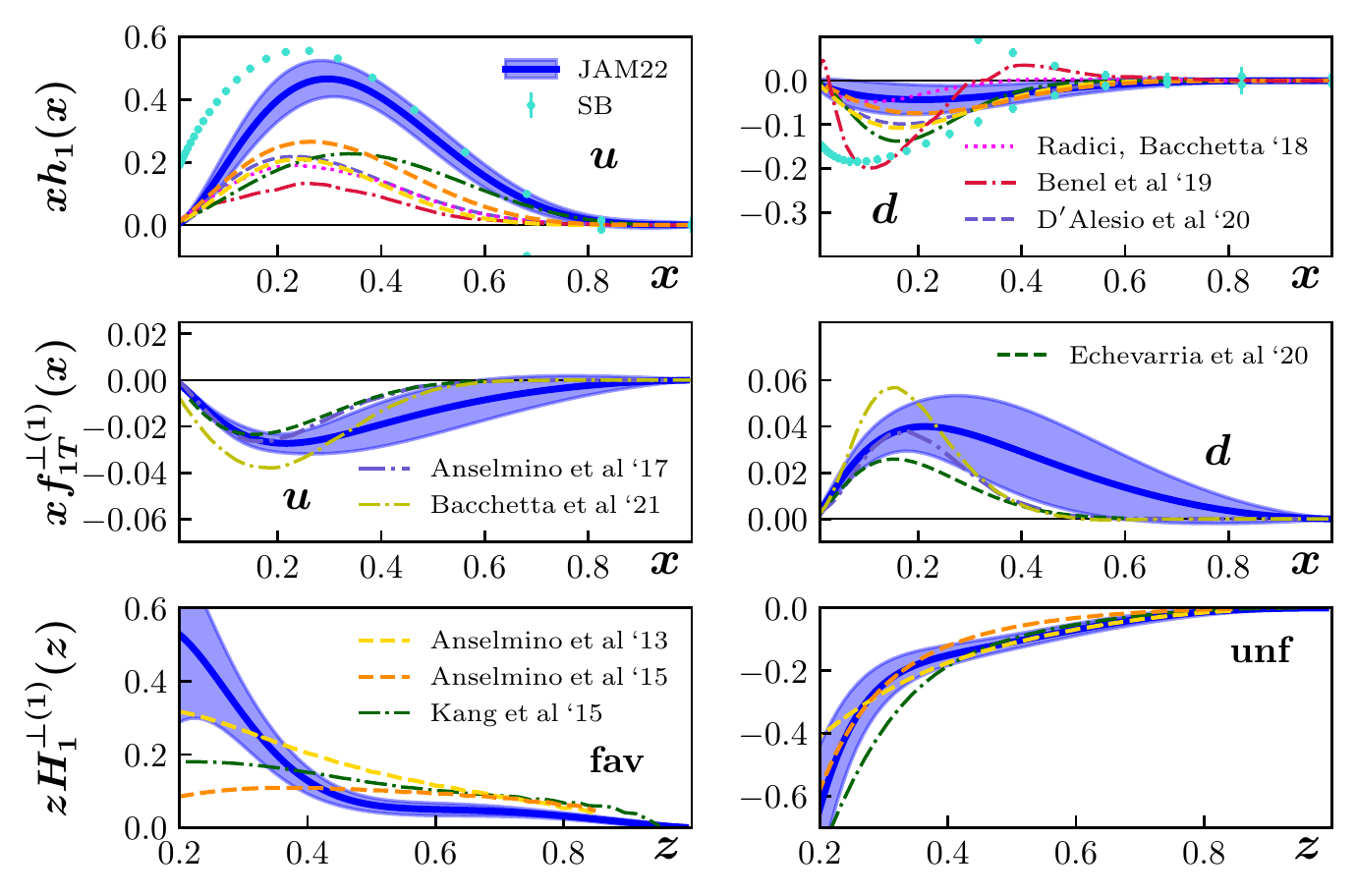}
\vspace{-0.2cm}
\caption{The extracted functions $h_1(x)$, $f_{1T}^{\perp(1)}(x)$, and $H_1^{\perp(1)}(z)$ at $Q^2=4$ GeV$^2$ from our {\tt JAM3D-22}  global analysis (blue solid curves with 1-$\sigma$ CL error bands) compared to the functions from other groups.  The generated Soffer bound (SB) data are also displayed (cyan points). We note that 
for all groups the curves are the central values of the $68\%$ confidence band. 
The transversity function for Radici, Bacchetta~`18 and Benel, et al.~'20 are for valence $u$ and $d$ quarks.
} 
\label{f:qcf_comparison}
\end{figure}
%%%%%%%%%%%%%%%%%

The comparison of our {\tt JAM3D-22}  non-perturbative functions with those from other groups is shown in Fig.~\ref{f:qcf_comparison}.  Now that the Soffer bound is imposed in {\tt JAM3D-22}, $h_1^d(x)$ matches more closely to other extractions than the {\tt JAM3D-20/20+} version displayed in Fig.~\ref{f:qcf}.  However, a striking difference is still the large size of $h_1^u(x)$ in {\tt JAM3D-22}  that now saturates the Soffer bound at $x\gtrsim 0.35$.  This is necessary to not only describe the lattice $g_T$ data point but also the $A_N^\pi$ measurements.  Without including this information in the analysis (i.e., relying only on the standard TMD or dihadron observables that are typically used to extract transversity), one does not find this solution for $h_1^u(x)$.  This function can actually describe all relevant SSAs considered here (TMD {\it and} collinear twist-3) sensitive to transversity as well as obtain agreement with lattice tensor charge values.  To further emphasize the fact that current TMD observables and lattice are compatible, we also re-ran our analysis including {\it only} TMD observables (SIDIS, SIA, DY), imposing the Soffer bound on transversity, and including the lattice $g_T$ data point.  We found, similar to Ref.~\cite{Lin:2017stx}, good agreement with experiment and lattice and a size for $h_1^u(x)$ that falls in between our {\tt JAM3D-22}  result and those from other groups ($h_1^d(x)$ remains similar to other groups, although slightly larger in magnitude than {\tt JAM3D-22}).  The remaining increase in $h_1^u(x)$ seen in {\tt JAM3D-22}  is due to the inclusion of $A_N^\pi$ data in the analysis. 

%%%%%%%%%%%%%%%%%%%%%%%%
\begin{figure}[h!]
\includegraphics[width=0.375\textwidth]{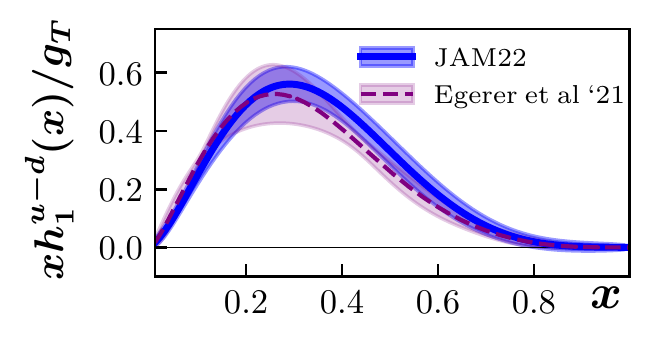}
\includegraphics[width=0.375\textwidth]{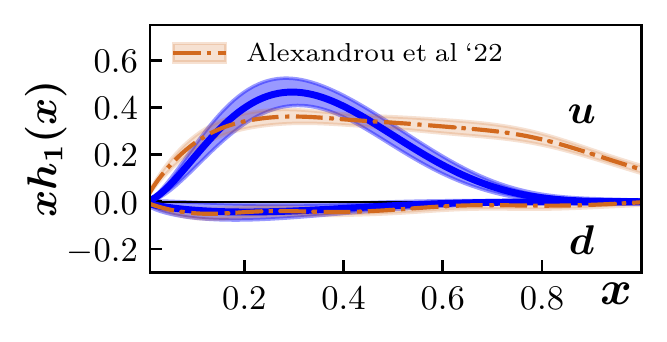}
\vspace{-0.2cm}
\caption{Plot of (left) $h_1^{u\text{-}d}(x)/g_T$, where $h_1^{ u\text{-}d}(x)\equiv h_1^u(x)-h_1^d(x)$, from the lattice calculation  of Ref.~\cite{HadStruc:2021qdf} (at $Q^2=2\,{\rm GeV^2}$) using $m_\pi= 358\,{\rm MeV}$ with statistical and systematic uncertainties added in quadrature (purple), and (right) $h_1^{u}(x)$ and $h_1^{d}(x)$ from the lattice calculation of Refs.~\cite{Alexandrou:2019lfo,Martha:privcom} (at $Q^2=4\,{\rm GeV^2}$) at the physical pion mass with only statistical uncertainties, compared to our {\tt JAM3D-22} result (blue) at $Q^2=4\,{\rm GeV^2}$.} \vspace{-0.3cm} 
\label{f:qcf_lat_comparison}
\end{figure}
%%%%%%%%%%%%%%%%%%%%%%%

Lattice QCD practitioners have also been able recently to calculate the $x$ dependence of transversity through the use of pseudo-PDFs~\cite{HadStruc:2021qdf} or quasi-PDFs~\cite{Alexandrou:2019lfo,Alexandrou:2021oih}.  The quantity extracted in Ref.~\cite{HadStruc:2021qdf} was $h_1^{ u\text{-}d}(x)/g_T$, where $h_1^{u\text{-}d}(x)\equiv h_1^u(x)-h_1^d(x)$, using $m_\pi= 358\,{\rm MeV}$.  Therefore, we plot this same combination in the left panel of Fig.~\ref{f:qcf_lat_comparison} and compare to the lattice result.  We find very good agreement across the entire $x$ range. The computation of $h_1^{u}(x)$ and $h_1^{d}(x)$ in Refs.~\cite{Alexandrou:2019lfo,Martha:privcom} was at the physical pion mass, and we compare {\tt JAM3D-22} to that result in the right panel of Fig.~\ref{f:qcf_lat_comparison}.  The agreement with $h_1^{u}(x)$ is good for $x\lesssim 0.5$. The difference in the large-$x$ region is
mostly due to systematic effects in the lattice results related to the reconstruction of the $x$ dependence from limited discretized data. This can be seen, for instance, in the comparison of different analyses of the same raw lattice data for the unpolarized case:~the quasi-PDF method~\cite{Alexandrou:2019lfo} has large-$x$ oscillations, whereas using a pseudo-PDF analysis alleviates the problem~\cite{Bhat:2020ktg}. Similarly, the pseudo-PDF analysis of Ref.~\cite{Joo:2020spy} has the expected decay of the unpolarized PDF in the large-$x$ region.  Notice that the lattice data presented in~\cite{Alexandrou:2019lfo,Martha:privcom} and in~\cite{HadStruc:2021qdf} are compatible with each other. As discussed, the discrepancy in the reconstructed shape is partly due to differences in the treatment of the lattice data in the quasi-PDF and pseudo-PDF approaches. Such a systematic effect is non-trivial to quantify. The agreement with $h_1^{d}(x)$ is very good for the entire $x$ range. 
Now that the lattice $g_T$ data point is included in {\tt JAM3D-22}, along with imposing the Soffer bound, we find the uncertainties in the phenomenological transversity function are similar to those from lattice QCD.

\begin{figure}[h!]
\includegraphics[width=0.775\textwidth]{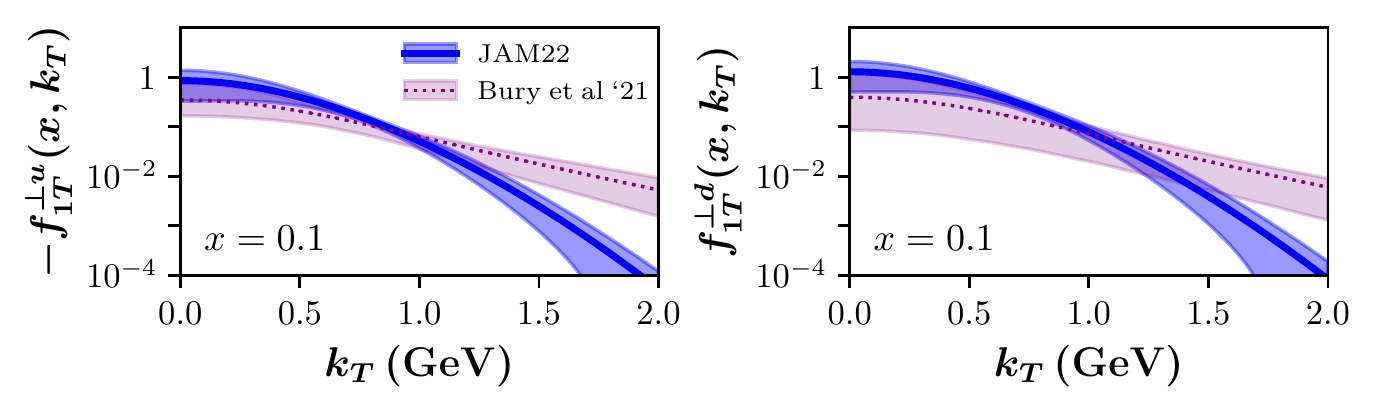}
\vspace{-0.3cm}
\caption{The extracted function $f_{1T}^{\perp}(x,k_T)$ at $x=0.1$ and at $Q^2=4\,{\rm GeV^2}$ as a function of $k_T$ (GeV) from our {\tt JAM3D-22} global analysis (blue solid curves with 1-$\sigma$ CL error bands) compared to the result from Ref.~\cite{Bury:2021sue}.}
\label{f:qcf_comparison_BPV}
\end{figure}

Lastly, the increase in size and slower fall off  at larger $x$ of $f_{1T}^{\perp(1)}(x)$ is a consequence of the 3D-binned HERMES Sivers effect data (see Appendix~\ref{s:app_a}).  This change in the function makes the magnitude of {\tt JAM3D-22}'s $f_{1T}^{\perp(1)}(x)$ more consistent with the recent extractions in Ref.~\cite{Echevarria:2020hpy} (Echevarria, et al.~`20 in Fig.~\ref{f:qcf_comparison}) as well as Ref.~\cite{Bacchetta:2020gko} (Bacchetta, et al.~`21 in Fig.~\ref{f:qcf_comparison}). However, in {\tt JAM3D-22} the fall off in the Sivers function at larger $x$ is generally slower than~\cite{Echevarria:2020hpy,Bacchetta:2020gko}.  We note that neither~\cite{Echevarria:2020hpy} nor~\cite{Bacchetta:2020gko} used the new 3D-binned HERMES data in their analyses. The method used in Ref.~\cite{Bury:2021sue} (Bury, et al.~`21 in Fig.~\ref{f:qcf_comparison_BPV}) to extract the Sivers function is different than the groups shown in Fig.~\ref{f:qcf_comparison}.  The authors directly extracted $\tilde{f}_{1T}^\perp(x,b_T)$, and the connection to the Qiu-Sterman function $F_{FT}(x,x)$ (and consequently $f_{1T}^{\perp(1)}(x)$) was made via a model independent inversion of the OPE relation at particular values of $Q=10$ GeV and $b_T= 0.11$ GeV$^{-1}$ that allow to minimize logarithmic corrections.  Therefore, in Fig.~\ref{f:qcf_comparison_BPV} we compare the Fourier transformed result of Ref.~\cite{Bury:2021sue} to our $k_T$-dependent function at $Q^2=4\, {\rm GeV^2}$.  
The curves are similar at small $k_T$ which suggests that at HERMES and COMPASS kinematics TMDs are predominantly dominated by non-perturbative contributions; however, they start to deviate from each other at larger values of $k_T$ due to the inclusion of gluon radiation effects in the analysis of Ref.~\cite{Bury:2021sue}. We expect that these effects will become important for the description of data from higher energies, such as the future EIC at larger $Q^2$. They also may become important for description of more precise STAR weak gauge boson data.

%%%%%%%%%%%%%%
%%%%%%%%%%%%%%

%%%%%%%%%%%%%%%%%%%%%%%%%%%%%%%%
%%%%%%%%%%%%%%%%%%%%%%%%%%%%%%%%
\subsection{Comparison between Theory and Experimental Data}\label{s:Discuss}
\begin{figure}[h]
\includegraphics[width=1\textwidth]{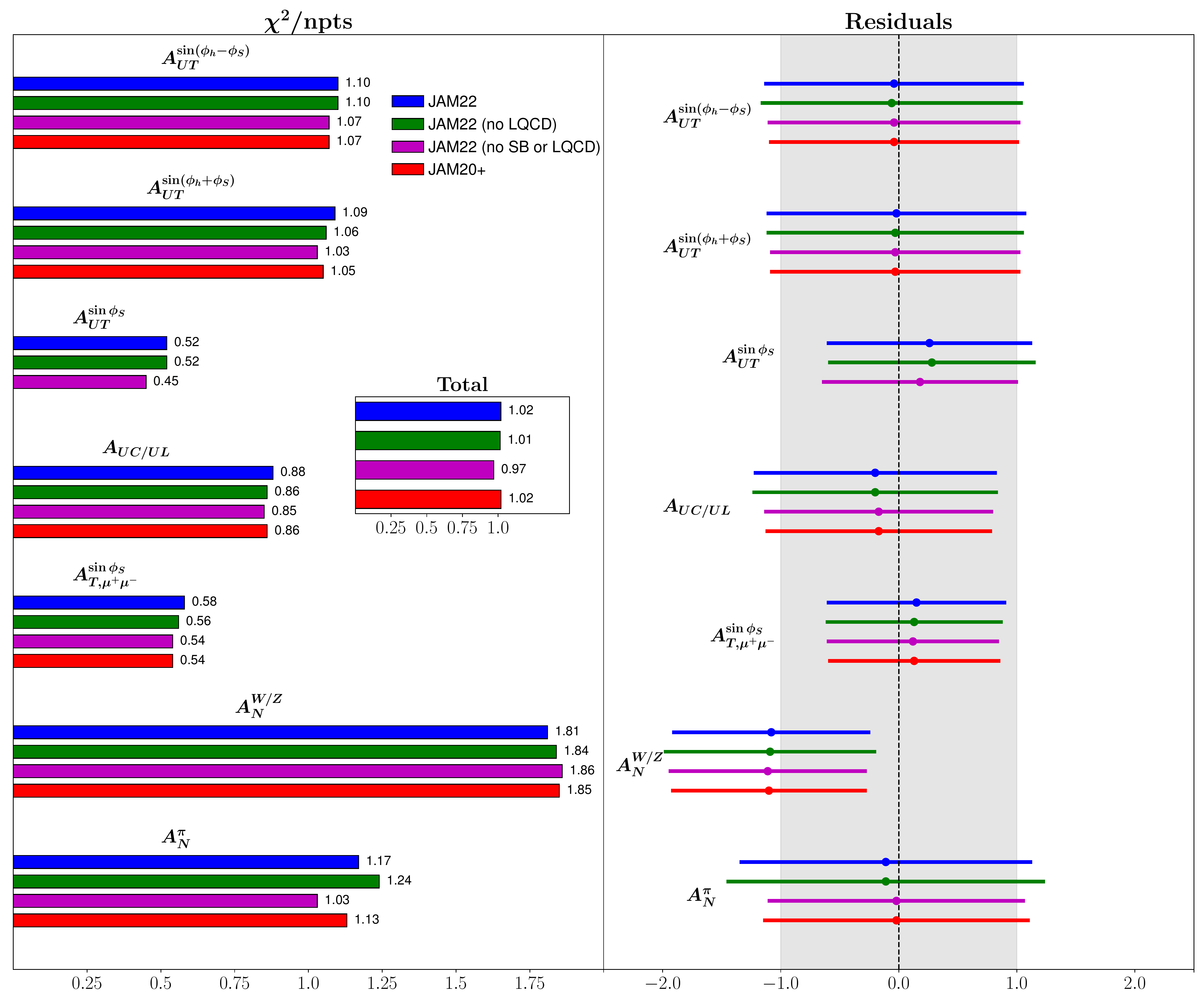}\vspace{-0.3cm}
\caption{The $\chi^2/{\rm npts}$ (left) and average value of the residuals with 1-$\sigma$ error bar (right) for each observable for various scenarios:~our full {\tt JAM3D-22}  global analysis (blue); {\tt JAM3D-22} without lattice QCD data (green); {\tt JAM3D-22} without both lattice QCD and SB data (purple); and {\tt JAM3D-20+} (red).  The inset in the left plot shows the total $\chi^2/{\rm npts}$.  Note that the {\tt JAM3D-20+} scenario does not include $A_{UT}^{\sin\phi_S}$ data, so no corresponding $\chi^2/{\rm npts}$ or residual is shown. \vspace{-0.25cm}} 
\label{f:chi2}
\end{figure}

%%%%%%%%%%%%%%%%%%%%%%%%%%%%%%%%
%%%%%%%%%%%%%%%%%%%%%%%%%%%%%%%%
In addition to the Sivers and Collins effects in SIDIS, Collins effect in SIA, Sivers effect in DY, and $A_N^\pi$ in proton-proton collisions that were a part of {\tt JAM3D-20/20+}, our {\tt JAM3D-22} global analysis of SSAs now includes the ($x$- and $z$-projected) $A_{UT}^{\sin\phi_S}$ asymmetry in SIDIS, a lattice QCD data point for $g_T$, and the Soffer bound constraint on transversity.  The comparison of our theory curves with the experimental data is shown in Appendix~\ref{s:app_b}.  The overall $\chi^2/{\rm npts}=1.02$, and the $\chi^2/{\rm npts}$ for the individual observables is also very good, as displayed in Table~\ref{t:sum} and Fig.~\ref{f:chi2}.

In order to systematically study the influence of the lattice QCD and Soffer bound (SB) constraints, we also performed an analysis where neither are included ({\tt JAM3D-22}  no SB or LQCD) as well as a study with SB data included but no lattice data on $g_T$ ({\tt JAM3D-22}  no LQCD).  A summary of the $\chi^2/{\rm npts}$ for the various scenarios is given in Fig.~\ref{f:chi2}. The quality of the results is very similar for each observable for all scenarios.
However, one can see that including the lattice $g_T$ data point and imposing the Soffer bound in the analysis play a noticeable role in the description of the data, in particular for $A_N^\pi$.  Recall that $A_{UT}^{\sin\phi_S}$ data is included in all the {\tt JAM3D-22}  scenarios.  This data is sensitive to the twist-3 FF $\tilde{H}(z)$ that also enters the fragmentation term for $A_N^\pi$ (see Eq.~(\ref{e:mathcalH})).  In {\tt JAM3D-20/20+}, $\tilde{H}(z)$ was set to zero. When $\tilde{H}(z)$ is now allowed to be nonzero, one finds initially an improvement in describing $A_N^\pi$.  Specifically, $\chi^2/{\rm npts}=1.03$ when neither the Soffer bound nor lattice $g_T$ data are included ({\tt JAM3D-22}  no SB or LQCD) compared to $1.13$ in {\tt JAM3D-20+}.  However, once there is a constraint from the Soffer bound ({\tt JAM3D-22}  no LQCD), the $\chi^2/{\rm npts}$ for $A_N^\pi$ increases to $1.23$.  Adding in the lattice $g_T$ data point (full {\tt JAM3D-22}) improves the situation for $A_N^\pi$ slightly to $\chi^2/{\rm npts}=1.17$.

The reason for the change in $\chi^2/{\rm npts}$ from {\tt JAM3D-20+} to ({\tt JAM3D-22}  no SB or LQCD) is due to the fact that the $\tilde{H}(z)$ term in $A_N^\pi$ gives a non-negligible contribution to the asymmetry that helps better describe the data, and $h_1(x)$ remains a similar size to {\tt JAM3D-20+}.  Once the Soffer bound is imposed ({\tt JAM3D-22}  no LQCD), this restricts the size of the transversity function, especially for the down quark, making the quality of the agreement with $A_N^\pi$ ($\chi^2/{\rm npts}=1.24$) not as good as ({\tt JAM3D-22}  no SB or LQCD) ($\chi^2/{\rm npts}=1.03$).  Including the lattice $g_T$ data point, which yields the full {\tt JAM3D-22}  result, causes a slight increase in the magnitude of $h_1(x)$ compared to ({\tt JAM3D-22}  no LQCD) (see Fig.~\ref{f:qcf_SB_LatgT}), thus providing a reasonable match to the data (see Fig.~\ref{f:an}) with a somewhat improved $\chi^2/{\rm npts}=1.17$.  The function $\tilde{H}(z)$ is key to maintaining agreement with $A_N^\pi$ measurements once the Soffer bound is imposed on transversity, as the $\chi^2/{\rm npts}$ for that observable increases to $1.53$ if $\tilde{H}(z)$ is kept fixed at zero.

The reason in {\tt JAM3D-22}  we are able to obtain a signal for $\tilde{H}(z)$ is due to the inclusion of the ($x$- and $z$-projected) $A_{UT}^{\sin\phi_S}$ asymmetry in SIDIS from HERMES, and we find a very good description to that data (see Fig.~\ref{f:sidis_sinphiS}), with $\chi^2/{\rm npts}=0.52$.  The HERMES and COMPASS data for the Collins and Sivers effects are in agreement with our analysis (see Figs.~\ref{f:sidis_HERMES_Collins}, \ref{f:sidis_HERMES_Sivers}, \ref{f:sidis_COMPASS}), with $\chi^2/{\rm npts}=1.09$ for the former and $\chi^2/{\rm npts}=1.10$ for the latter.  The Collins SIA data is also described well (see Fig.~\ref{f:sia}) with $\chi^2/{\rm npts}=0.88$.  The pion-induced DY data for $\mu^+\mu^-$ pairs from COMPASS and proton-proton $W$ and $Z$ production from STAR compared to our theory curves are shown in Fig.~\ref{f:DY}.  The former matches our theory calculation ($\chi^2/{\rm npts}=0.58$) while the latter has some discrepancies ($\chi^2/{\rm npts}=1.81$).  New preliminary data from STAR~\cite{Eyser:talkSPIN21}, though, shows noticeable changes to what is displayed in Fig.~\ref{f:DY} that should bring theory and experiment in closer alignment.  Nevertheless, we explored using a different input for the unpolarized proton and pion PDFs but did not find any remarkable change to the $\chi^2/{\rm npts}$. Since the $x$ coverage (in the transversely polarized proton) for both COMPASS and STAR is similar ($0.1\lesssim x \lesssim 0.35)$ while the $Q^2$ values are much different ($Q^2\approx 30\,{\rm GeV^2}$ for COMPASS and $\approx(80\,{\rm GeV})^2$ for STAR), one may think the higher $\chi^2/{\rm npts}$ for the latter is a consequence of how we evolve the Sivers function. However, the analyses in Refs.~\cite{Echevarria:2020hpy,Bacchetta:2020gko} had similar issues with the STAR DY data, even with including TMD evolution. 

%%%%%%%%%%%%%%%%%%%%%%%%%%%%%%%%%%%%%%%%%%%%%%%%%%%%%%%%%%%%%%%%%%
\section{Exploratory Study on the Role of Antiquarks}\label{s:Antiquarks}  
\begin{figure}[h!]
\includegraphics[width=0.7\textwidth]{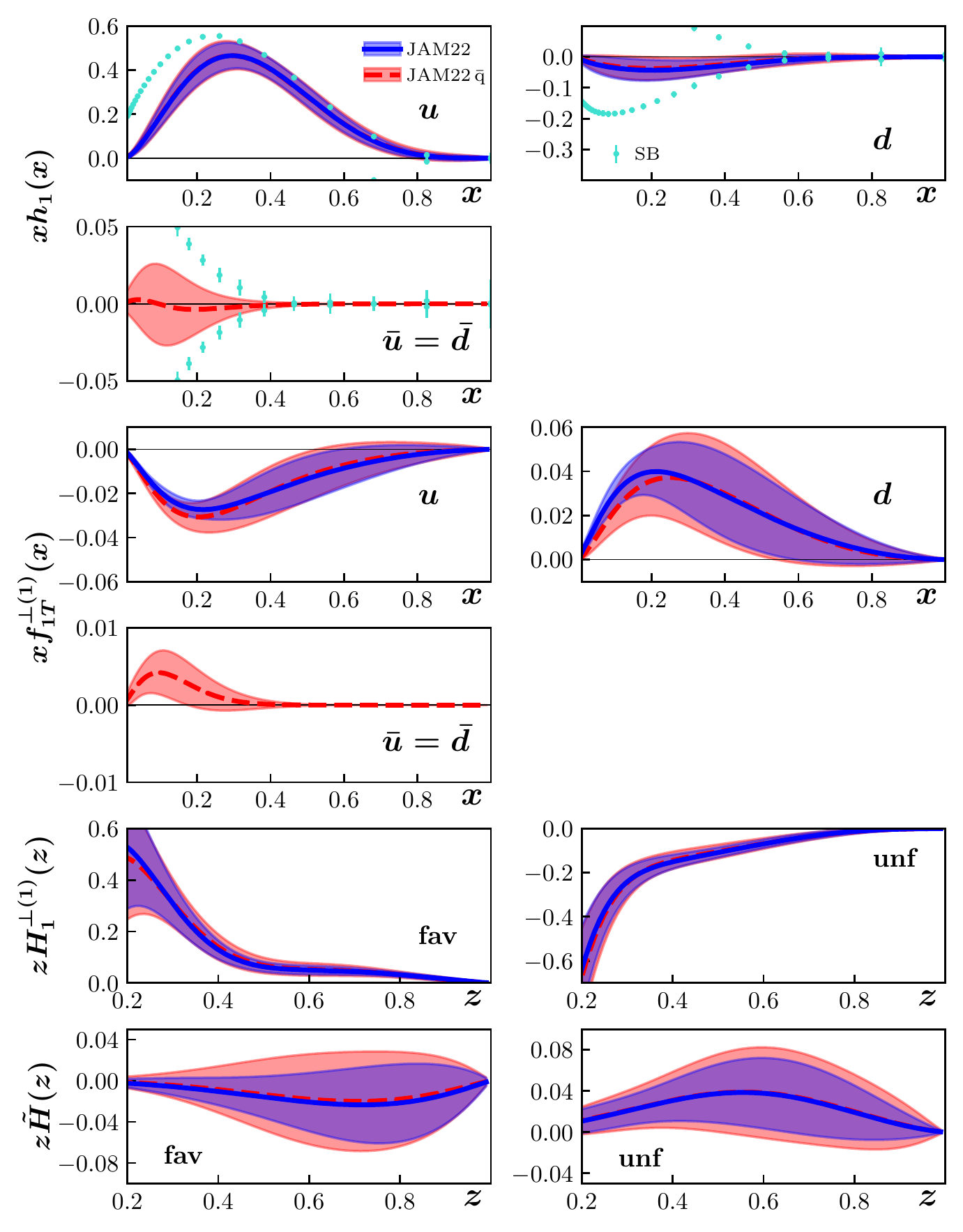}
\vspace{-0.2cm}
\caption{The extracted non-perturbative functions  for {\tt JAM3D-22} (blue) from Sec.~\ref{s:Pheno} (no antiquarks), and the analysis {\tt JAM3D-22$\,{\rm \bar{q}}$} including antiquarks (red), where $\bar{u}=\bar{d}$.} 
\label{f:qcf_JAM22qbar}
\end{figure}

\begin{figure}[h!]
\includegraphics[width=0.65\textwidth]{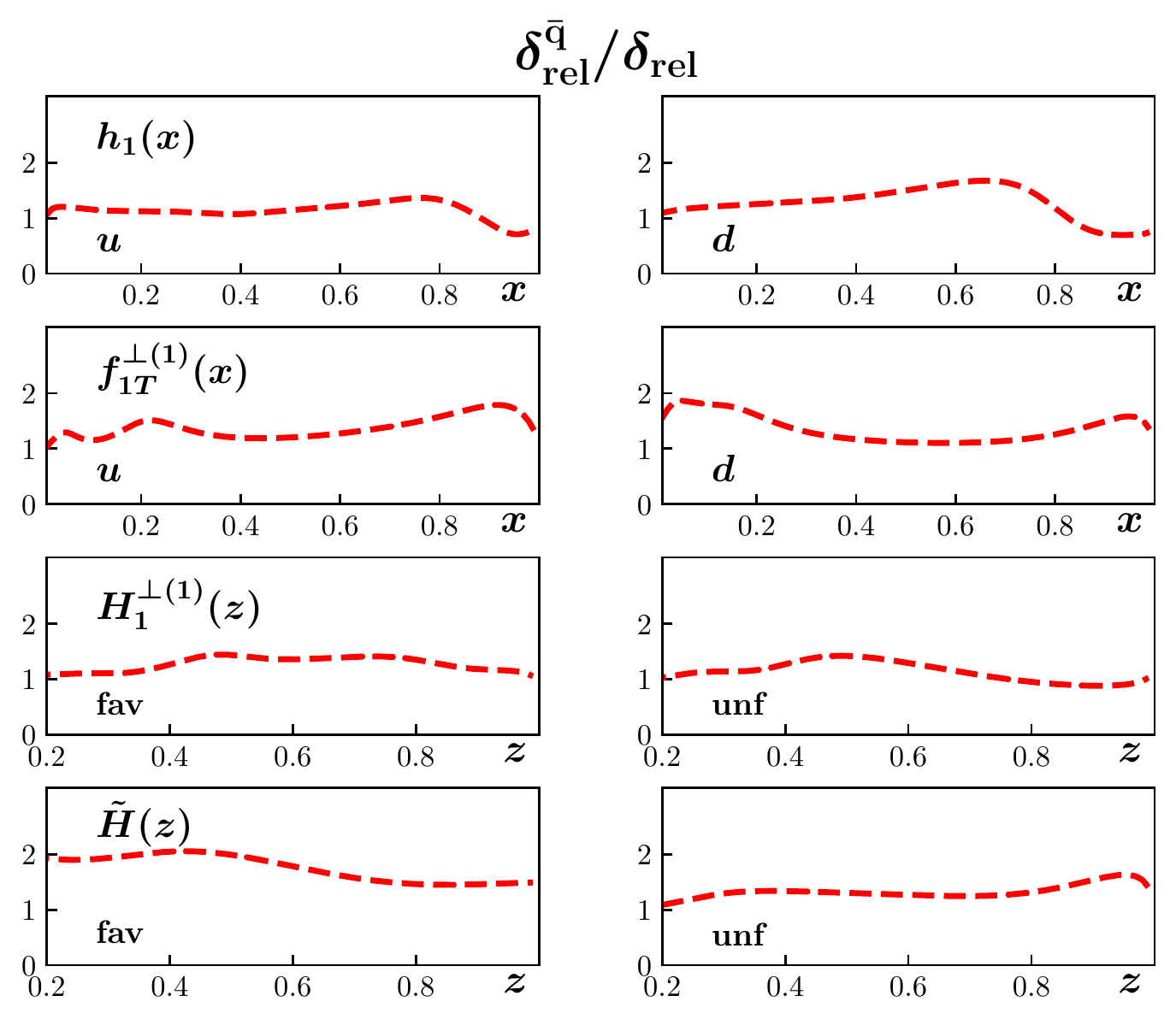}
\vspace{-0.2cm}
\caption{The ratio of the relative uncertainty $\delta_{\rm rel}^{\bar{q}}$ for the analysis {\tt JAM3D-22$\,{\rm \bar{q}}$} including antiquarks to the relative uncertainty $\delta_{\rm rel}$ of {\tt JAM3D-22} from Sec.~\ref{s:Pheno} (no antiquarks).} 
\label{f:relratio}
\end{figure}

%%%%%%%%%%%%%%%%%%%%%%%%%%%%%%%%%%%%%%%%%%%%%%%%%%%%%%%%%%%%%%%%%%
\vspace{-0.15cm}
The analysis we presented so far in this paper follows the construction of {\tt JAM3D-20}, which did not include antiquarks. Indeed, previous phenomenological analyses found either small or negligible contributions from antiquarks~\cite{Echevarria:2014xaa,Kang:2015msa,Anselmino:2016uie,Echevarria:2020hpy,Bacchetta:2020gko}. Lattice QCD studies also suggest that the transversity antiquark functions are small~\cite{Alexandrou:2021oih}. However, for future facilities, such as the EIC, extracting antiquark contributions will become more important.
The inclusion of antiquarks may also help in describing the $A_N^{W/Z}$ data, as these terms can play more of a role in proton-proton collisions (likewise for $A_N^\pi$). Therefore, we perform a preliminary analysis ({\tt JAM3D-22$\,{\rm \bar{q}}$}) that includes $\bar{u}$ and $\bar{d}$ for the transversity and Sivers functions.  We assume a symmetric sea ($\bar{u}=\bar{d}$) with $N_{\bar{u}},\alpha_{\bar{u}},\beta_{\bar{u}}$ free parameters for each respective function.  The transverse momentum widths are taken to be the same as their quark counterparts.  The results are shown in Fig.~\ref{f:qcf_JAM22qbar} compared to the {\tt JAM3D-22} analysis from Sec.~\ref{s:Pheno}.  The antiquark functions are about $10-20\%$ of the quark ones, peak at smaller $x$, and fall off rapidly at larger $x$.  
As expected, there is an increase in the uncertainty of all the non-perturbative functions.  We quantify this by calculating the ratio of the relative uncertainties between {\tt  JAM3D-22$\,{\rm \bar{q}}$} and {\tt JAM3D-22}, which are shown in Fig.~\ref{f:relratio}.  We find that this ratio is close to 1 in most regions of $x/z$ but may become as large as 2, especially for $\tilde{H}(z)$.
%%%%%%%%%%%%%%%%%%%%%%%%%%%%%%%%%%%
%%%%%%%%%%%%%%%%%%%%%%%%%%%%%%%%%%%
%%%%%%%%%%%%%%%%%%%%%%%%%%%%%%%%%%%

\begin{figure}[h!]
\includegraphics[width=1\textwidth]{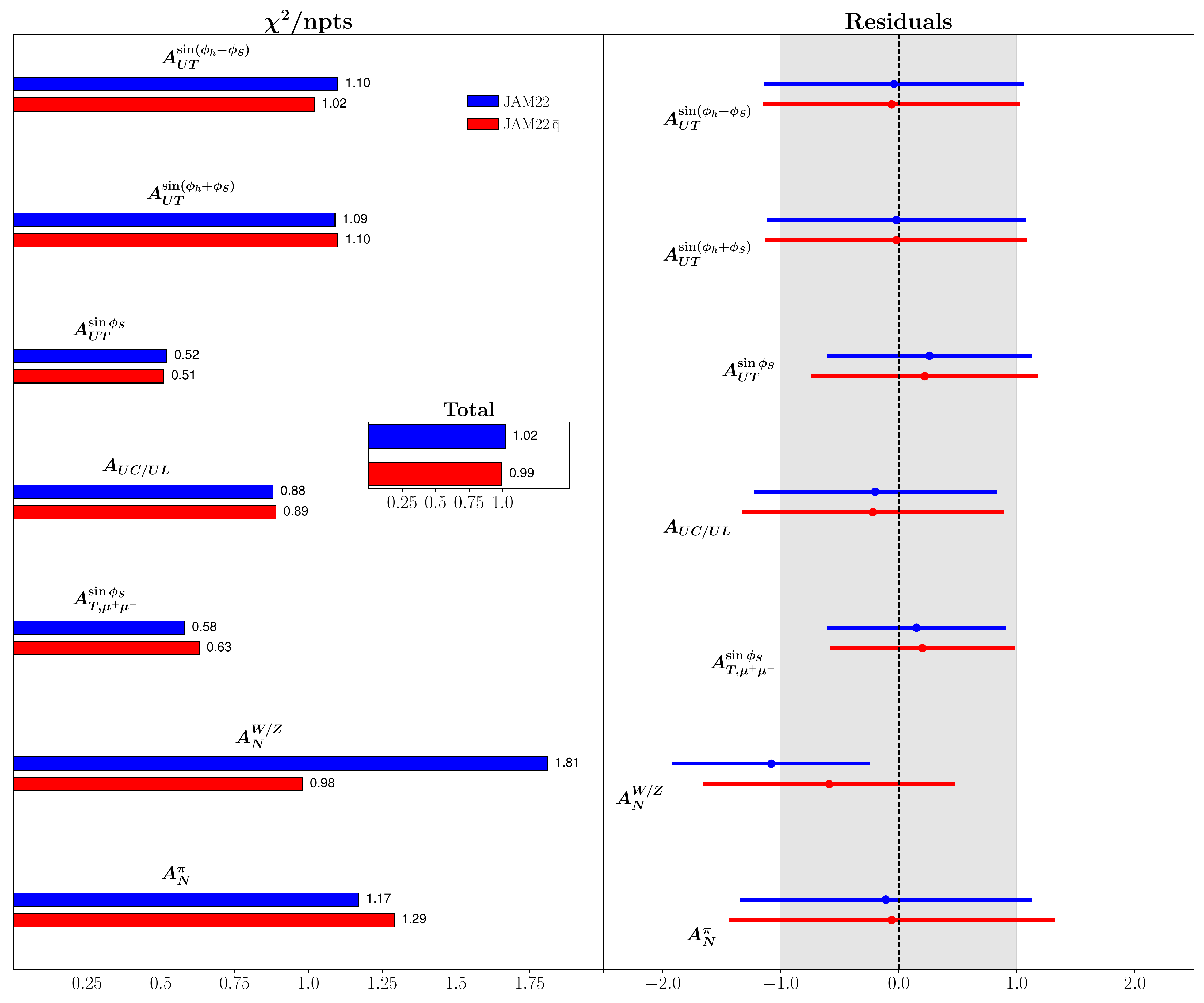}\vspace{-0.3cm}
\caption{The $\chi^2/{\rm npts}$ (left) and average value of the residuals with 1-$\sigma$ error bar (right) for each observable for  {\tt JAM3D-22} (blue) from Sec.~\ref{s:Pheno} (no antiquarks), and the analysis {\tt JAM3D-22$\,{\rm \bar{q}}$} including antiquarks (red). \vspace{-0.25cm}} 
\label{f:chi2_qbar}
\end{figure}

To see the impact on the observables, we display the $\chi^2/{\rm npts}$ for {\tt JAM3D-22$\,{\rm \bar{q}}$} and {\tt JAM3D-22} in Fig.~\ref{f:chi2_qbar}. $A_N^{W/Z}$ had the greatest improvement in its $\chi^2/{\rm npts}$ (going from 1.81 down to 0.98), with $A_{T,\mu^+\mu^-}^{\sin\phi_S}$ and $A_N$ becoming slightly worse.  However, all residuals still significantly overlap within their 1-$\sigma$ error, indicating that there is not a statistically significant change by including antiquarks. 
%%%%%%%%%%%%%%%%%%%%%%%%%%%%%%%
%%%%%%%%%%%%%%%%%%%%%%%%%%%%%%
We also confirm that the tensor charges $\delta u, \delta d, g_T$ are basically unchanged between {\tt JAM3D-22} and {\tt JAM3D-22$\,{\rm \bar{q}}$}.  In {\tt JAM3D-22} of Sec.~\ref{s:Pheno} we have $\delta u=0.78\pm 0.11, \delta d=-0.12\pm 0.11$, and $g_T=0.90\pm 0.05$, whereas in {\tt JAM3D-22$\,{\rm \bar{q}}$} we find $\delta u=0.78\pm 0.10, \delta d=-0.11\pm 0.10$, and $g_T=0.89\pm 0.08$. 
In general, the main conclusions of the paper remain unchanged.  We plan to include additional proton-proton data/observables from RHIC in our second update and also anticipate finalized (more precise) measurements of $A_N^{W/Z}$ forthcoming from STAR.  Therefore, we leave a more rigorous study on the impact of antiquarks to this future analysis.

%%%%%%%%%%%%%%%%%%%%%%%%%%%%%%%%%%%%%%%%%%%%%%%%%%%%%%%%%%%%%%%%%%
\section{Summary and Outlook}\label{s:Concl}  
%%%%%%%%%%%%%%%%%%%%%%%%%%%%%%%%%%%%%%%%%%%%%%%%%%%%%%%%%%%%%%%%%%
We have presented the first of two updates to the {\tt JAM3D-20} global analysis of SSAs~\cite{Cammarota:2020qcw}.  We include the same experimental data from that work, except for the following changes:~(i) HERMES data is replaced with their superseding 3D-binned measurements~\cite{HERMES:2020ifk}; (ii) data on $A_{UT}^{\sin\phi_S}$~\cite{HERMES:2020ifk} is included; (iii) the Soffer bound on transversity is imposed; (iv) the lattice tensor charge $g_T$ value from Ref.~\cite{Alexandrou:2019brg} is now included as a data point.  Before extending our analysis through (ii)--(iv), we first carried out (i) to form a new baseline ({\tt JAM3D-20+}).  The only noticeable change was to the Sivers function (see Appendix~\ref{s:app_a}), which, for both the up and down quarks, is now larger in magnitude and falls off slower at larger $x$, although the relative uncertainty remains the same.

With regard to our full updated result ({\tt JAM3D-22}), unlike {\tt JAM3D-20/20+}, we used the complete analytical result for the $A_N^\pi$ fragmentation term that includes the twist-3 FF $\tilde{H}(z)$ coupled to the transversity function $h_1(x)$.  These functions also enter the $\vec{P}_{hT}$-integrated $A_{UT}^{\sin\phi_S}$ asymmetry in SIDIS (see Eqs.~(\ref{e:SIDIS_SSAs}), (\ref{e:sinphiS})).   Using the $x$- and $z$-projected data on $A_{UT}^{\sin\phi_S}$ from HERMES~\cite{HERMES:2020ifk} as a proxy for the $\vec{P}_{hT}$-integrated observable, along with $A_N^\pi$ measurements from BRAHMS and STAR, we were able to obtain  the first extraction of $\tilde{H}(z)$ from a global analysis.  We found the function behaves similar to the Collins FF, where favored and unfavored are opposite in sign and roughly equal in magnitude.  This is consistent with previous information in the literature~\cite{Kanazawa:2014dca,Lu:2015wja}. We found that including the $\tilde{H}(z)$ term in $A_N^\pi$ improves agreement with that data compared to {\tt JAM3D-20}.

We also studied the impact that the Soffer bound and lattice $g_T$ data have on our extracted non-perturbative functions.  The Soffer bound causes a significant change to down quark transversity function compared to {\tt JAM3D-20/20+}, while the up quark is also slightly altered (see Fig.~\ref{f:qcf}).  Nevertheless, one is still able to maintain good agreement with all experimental data sets (see Fig.~\ref{f:chi2}).  Including the lattice data point on $g_T$ causes about a $50\%$ reduction in the uncertainty for $h_1(x)$ and also slightly decreases the uncertainty in $\tilde{H}(z)$ (see Fig.~\ref{f:qcf_SB_LatgT}).  

The {\tt JAM3D-20/20+} global analysis was the first time that phenomenology agreed with lattice for all the nucleon tensor charges $\delta u$, $\delta d$, and $g_T$.  The cause of this was the inclusion of $A_N^\pi$ data, which increased the magnitude of $h_1^u(x)$ and $h_1^d(x)$ compared to analyses that only included TMD or dihadron observables.  With the $\tilde{H}(z)$ term of $A_N^\pi$ now nonzero for {\tt JAM3D-22}, the transversity functions do not need to be as large, and one consequently finds that the tensor charge values, when one does not include the lattice $g_T$ data point ({\tt JAM3D-22}  no LQCD) fall below the lattice results (see Fig.~\ref{f:gT}).  However, the full {\tt JAM3D-22} that includes the lattice $g_T$ data point in the analysis is able to find agreement again with $\delta u$, $\delta d$, and $g_T$ along with all experimental data.  This highlights that fact, also concluded in Ref.~\cite{Lin:2017stx}, that one should include lattice constraints in an analysis in order to definitively determine if a tension exists with experimental data.  In the end, we find that our updated global analysis is able to accommodate the constraints from the Soffer bound and lattice QCD while also describing all experimental data from {\tt JAM3D-20/20+}.  The function $\tilde{H}(z)$ played a key role in maintaining agreement with the $A_N^\pi$ measurements.  
We created a user-friendly jupyter notebook accessible via {\tt Google Colab}~\cite{jam3dlib} that allows one to access  all the functions and asymmetries from our analysis.

As a second  update to our global analysis, we  will study the impact of new STAR data on $A_N^{\pi^0}$~\cite{STAR:2020nnl} as well as A$_N$DY and STAR data on $A_N^{jet}$~\cite{Bland:2013pkt,STAR:2020nnl} and STAR measurements of the Collins effect in hadron-in-jet for $\pi^\pm$~\cite{STAR:2017akg} and $\pi^0$~\cite{STAR:2020nnl}. This will allow for a comprehensive assessment of whether SSAs in both TMD and CT3 processes can be described in a unified framework and have a common origin.  The effects of TMD evolution will be systematically incorporated in future work.

\section*{Acknowledgments}  
We thank Martha Constantinou for many useful discussions about lattice QCD calculations. This work has been supported by the National Science Foundation under Grants No.~PHY-2011763 (MM, JAM, DP) and No.~PHY-2012002 (AP), the U.S. Department of Energy under contracts No. DE-FG02-07ER41460 (LG), No.~DE-AC05-06OR23177 (AP, NS) under  which  Jefferson  Science  Associates,  LLC,  manages and operates Jefferson Lab, and within the framework of the TMD Topical Collaboration.  The work of NS was also supported by the DOE, Office of Science, Office of Nuclear Physics in the Early Career Program.

%%%%%%%%%%%%%%%%%%%%%%%%%%%%%%%%%%%%%%%%%%%%%%%%%%%%%%%%%%%%%%%%%%
\appendix
%%%%%%%%%%%%%%%%%%%%%%%%%%%%%%%%%%%%%%%%%%%%%%%%%%%%%%%%%%%%%%%%%%
\section{Using 3D-Binned HERMES Data on the Sivers and Collins Effects}
\label{s:app_a}
\begin{figure}[h!]
\centering
\includegraphics[width=0.7\textwidth]{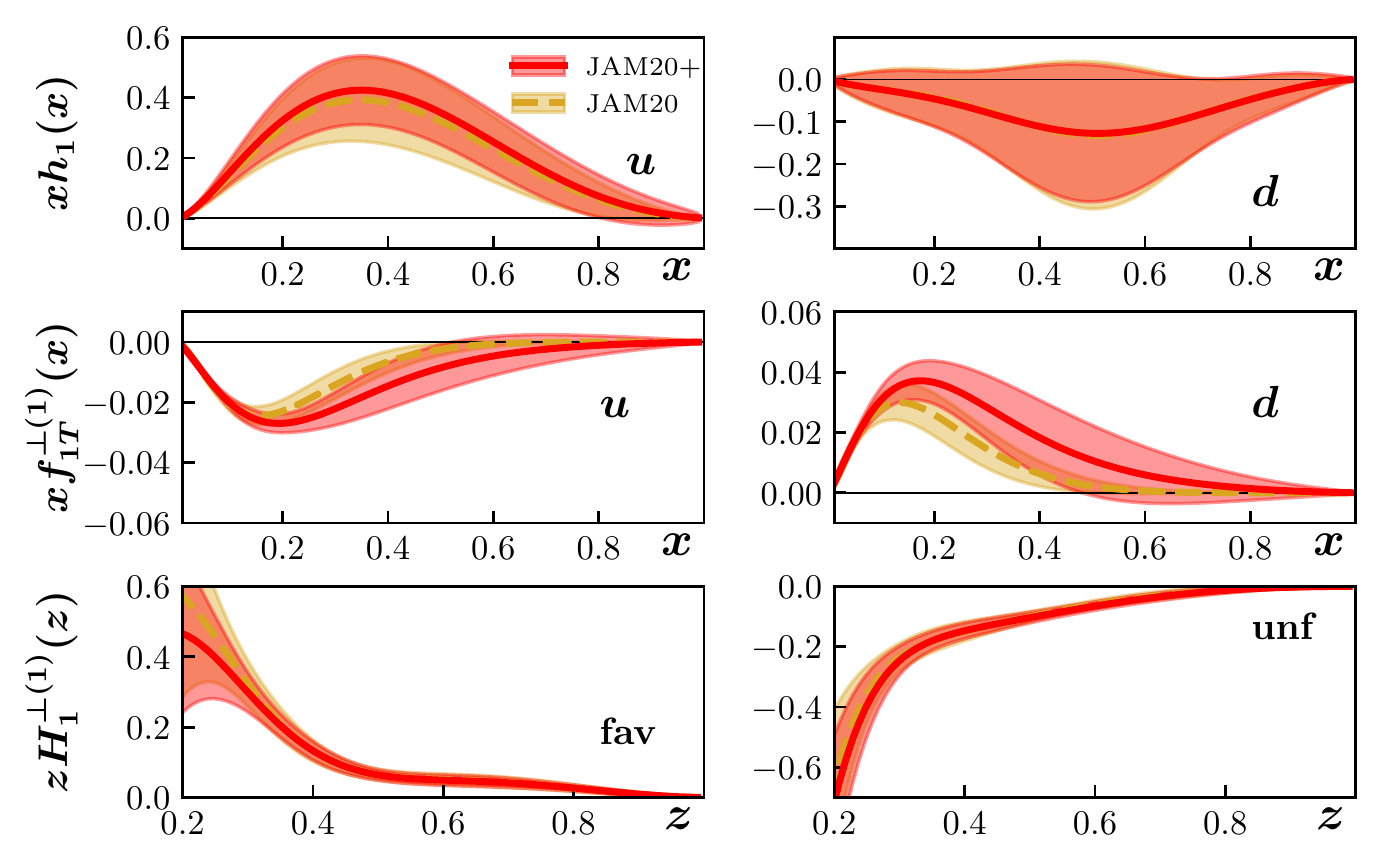}\vspace{-0.3cm}
\caption{
The extracted functions $h_1(x)$, $f_{1T}^{\perp(1)}(x)$, and $H_1^{\perp(1)}(z)$ at $Q^2=4$ GeV$^2$ from the {\tt JAM3D-20} global analysis~\cite{Cammarota:2020qcw} (light brown dashed curves with 1-$\sigma$ CL error bands) compared to {\tt JAM3D-20+} (red solid curves with 1-$\sigma$ CL error bands) that replaces the older HERMES data on the Sivers and Collins effects~\cite{Airapetian:2009ae,Airapetian:2010ds} (used in {\tt JAM3D-20}) by their new, 3D-binned measurements~\cite{HERMES:2020ifk}.} 
\label{f:qcf_JAM20plus}
\end{figure}

\begin{figure}[h!]
\centering
\includegraphics[width=0.7\textwidth]{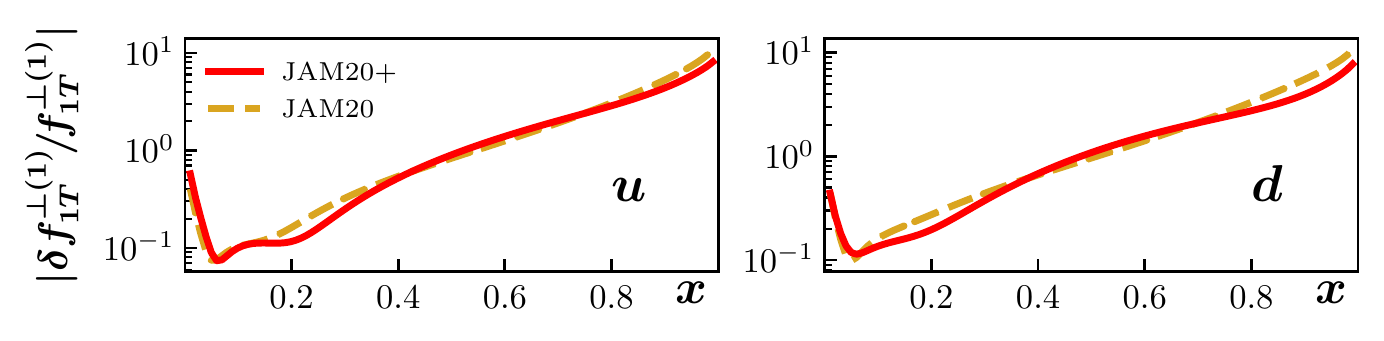}\vspace{-0.5cm}
\caption{
The relative error of $f_{1T}^{\perp(1)}(x)$ for {\tt JAM3D-20} (light brown dashed curve) and {\tt JAM3D-20+} (red solid curve).} 
\label{f:qcf_JAM20plus_Sivers_relerr}
\end{figure}

%%%%%%%%%%%%%%%%%%%%%%%%%%%%%%%%
%%%%%%%%%%%%%%%%%%%%%%%%%%%%%%%%
In this appendix we show the changes to the {\tt JAM3D-20} functions that occur when one replaces the HERMES data on the Sivers and Collins effects~\cite{Airapetian:2009ae,Airapetian:2010ds} with their new, 3D-binned data~\cite{HERMES:2020ifk} -- see Fig.~\ref{f:qcf_JAM20plus}.  We call this analysis {\tt JAM3D-20+}  and use it as our baseline to which we compare our {\tt JAM3D-22}  results in Fig.~\ref{f:qcf}. The only noticeable change is that the Sivers function becomes larger and falls off more slowly at larger $x$. In Fig.~\ref{f:qcf_JAM20plus_Sivers_relerr} we plot the relative uncertainty $\delta f_{1T}^{\perp(1)}(x)/f_{1T}^{\perp(1)}(x)$ for the Sivers first moment for both {\tt JAM3D-20} and {\tt JAM3D-20+}.  
Even though the {\tt JAM3D-20+}  Sivers function in Fig.~\ref{f:qcf_JAM20plus} may seem to have a larger error band than {\tt JAM3D-20}, we see in fact this is not the case.  Both the {\tt JAM3D-20} and {\tt JAM3D-20+}  functions have similar relative uncertainties.
%%%%%%%%%%%%%%%%%%%%%%%%
%%%%%%%%%%%%%%%%%%%%%%%
%%%%%%%%%%%%%%%%%%%%%%%%%
%%%%%%%%%%%%%%%%%%%%%%%%

%%%%%%%%%%%%%%%%%%%%%%%%%%%%%%%%
%%%%%%%%%%%%%%%%%%%%%%%%%%%%%%%%
\section{Comparison of Theory Results to Experimental Data}
\label{s:app_b}
\begin{figure}[h!]
\centering
\includegraphics[width=0.8\textwidth]{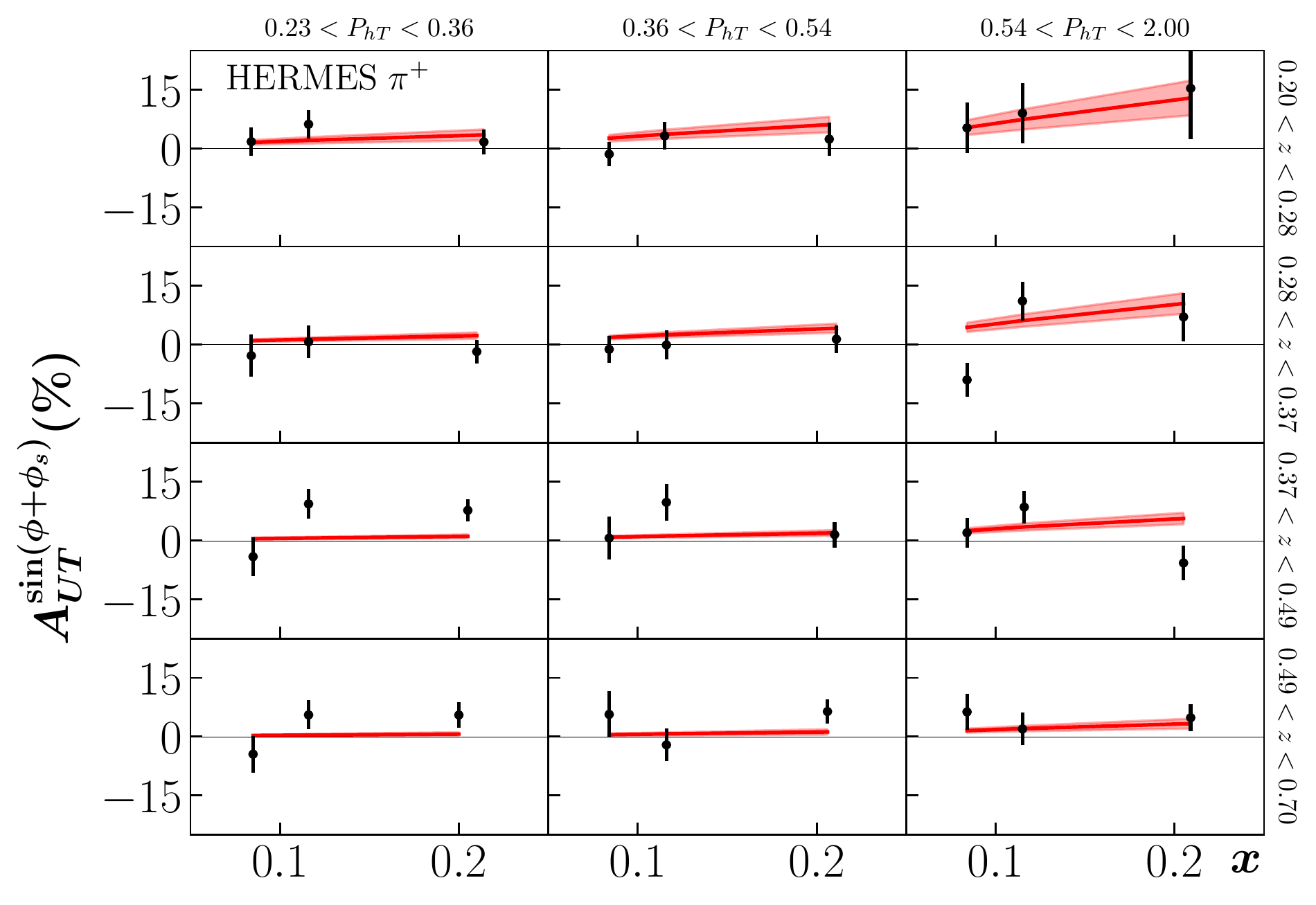}
\includegraphics[width=0.8\textwidth]{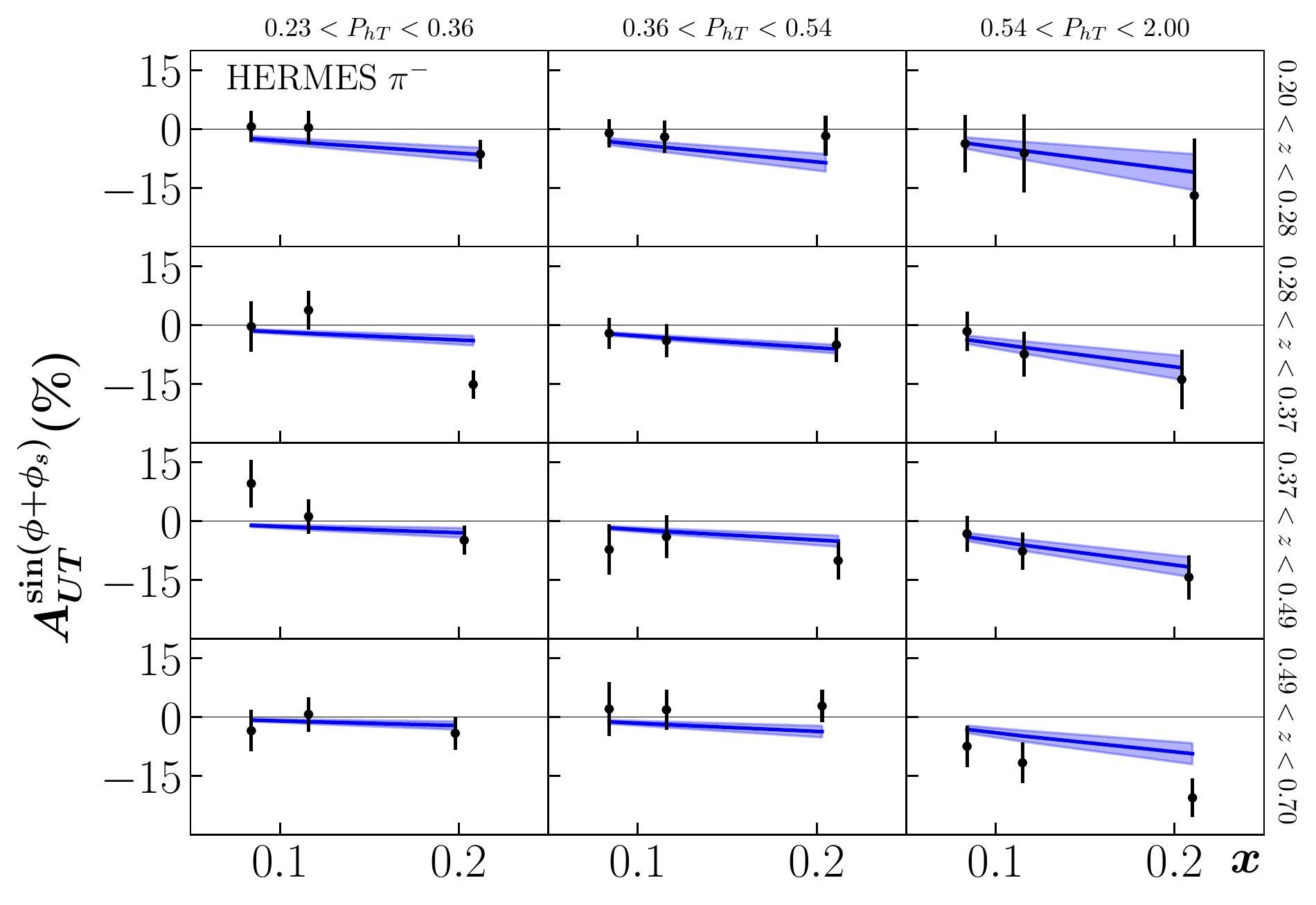}
\vspace{-0.3cm}
\caption{Plot of the Collins $A_{UT}^{\sin(\phi_h+\phi_S)}$ asymmetry for $\pi^+$ (top) and $\pi^-$ (bottom) for the 3D-binned HERMES data compared to our {\tt JAM3D-22}  global analysis.
} 
\label{f:sidis_HERMES_Collins}
\end{figure}

\begin{figure}[h!]
\centering
\includegraphics[width=0.8\textwidth]{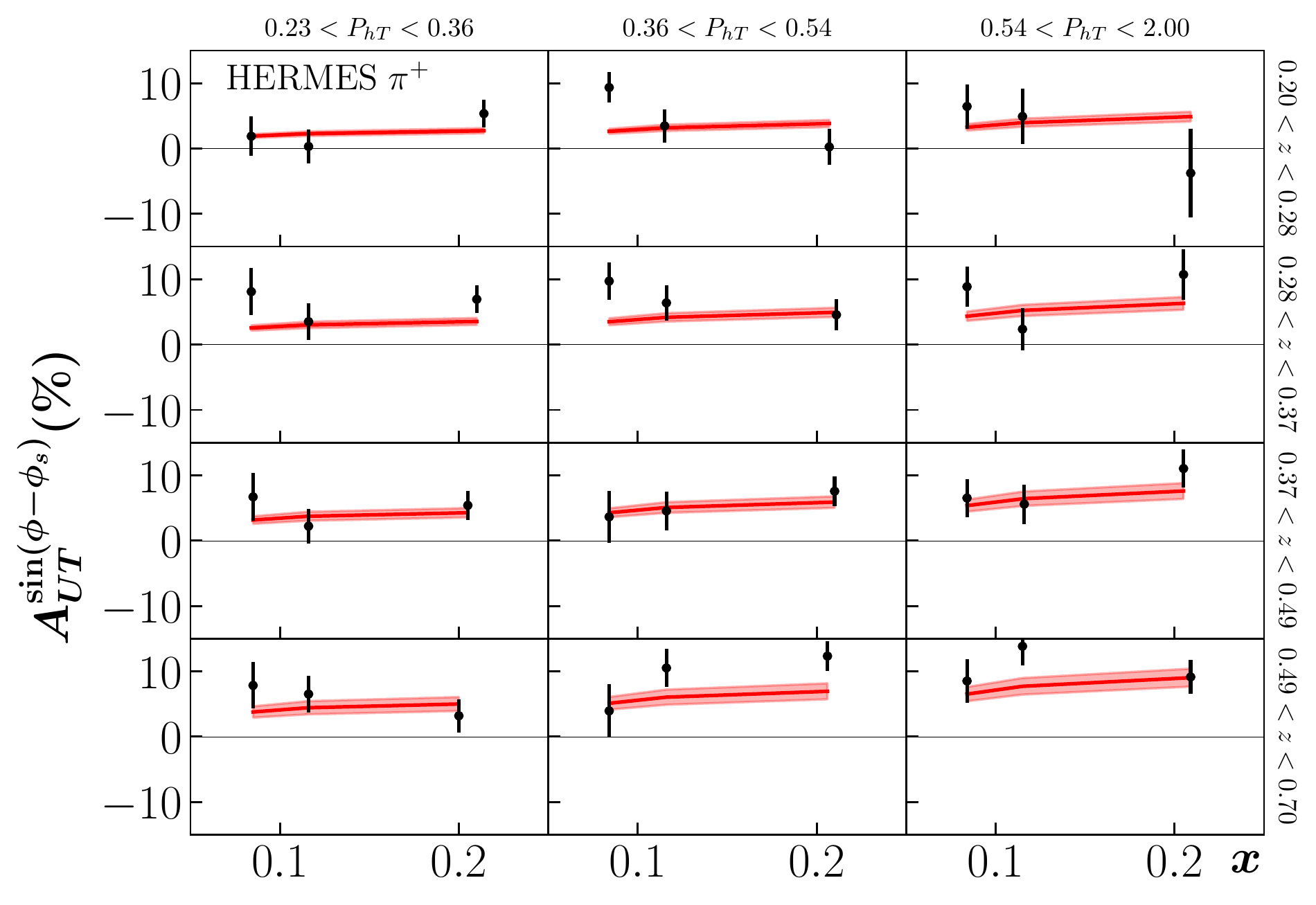}
\includegraphics[width=0.8\textwidth]{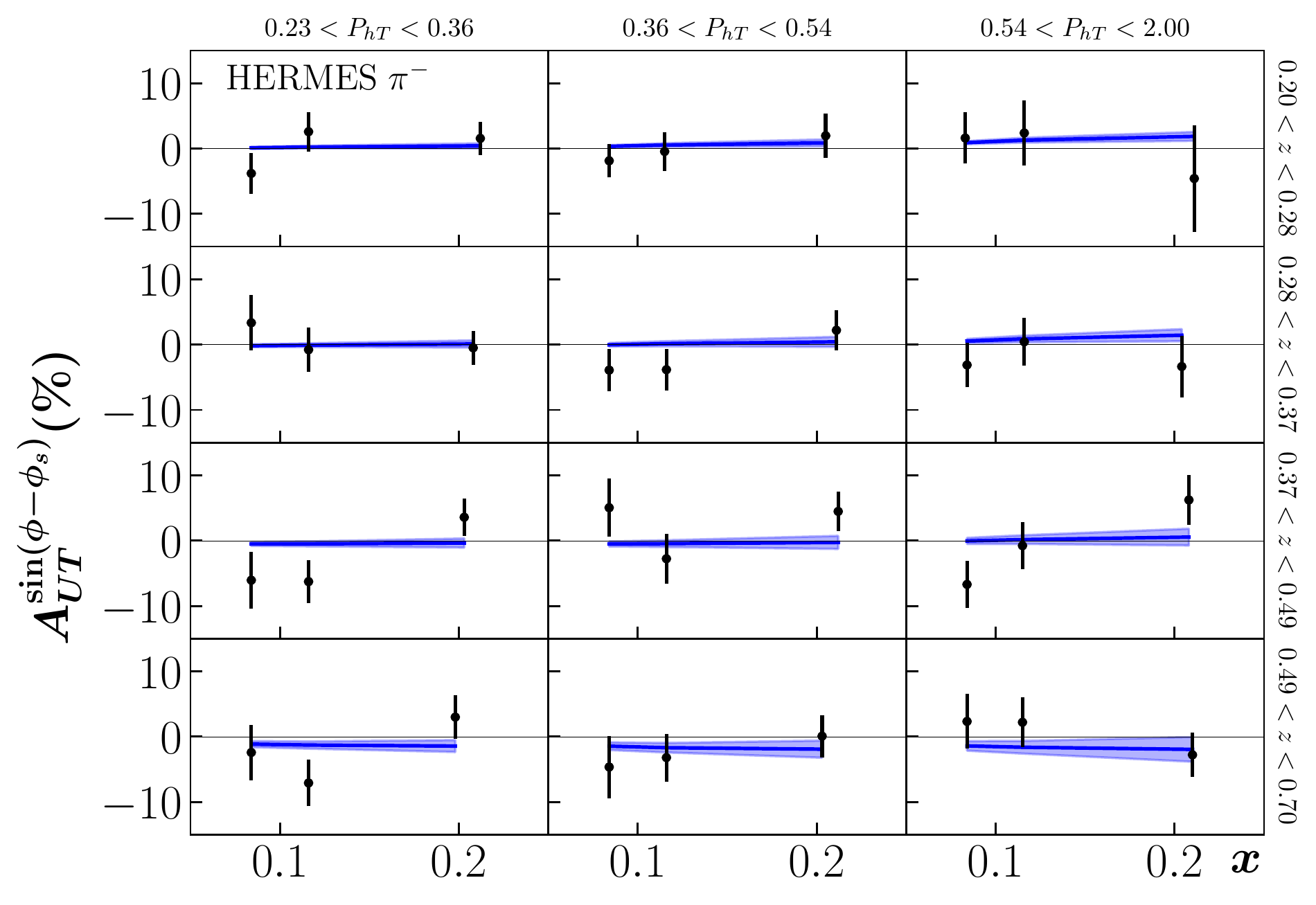}
\vspace{-0.3cm}
\caption{Plot of the Sivers $A_{UT}^{\sin(\phi_h-\phi_S)}$ asymmetry for $\pi^+$ (top) and $\pi^-$ (bottom) for the 3D-binned HERMES data  compared to our {\tt JAM3D-22}  global analysis.
} 
\label{f:sidis_HERMES_Sivers}
\end{figure}

\begin{figure}[h!]
\centering
\includegraphics[width=0.75\textwidth]{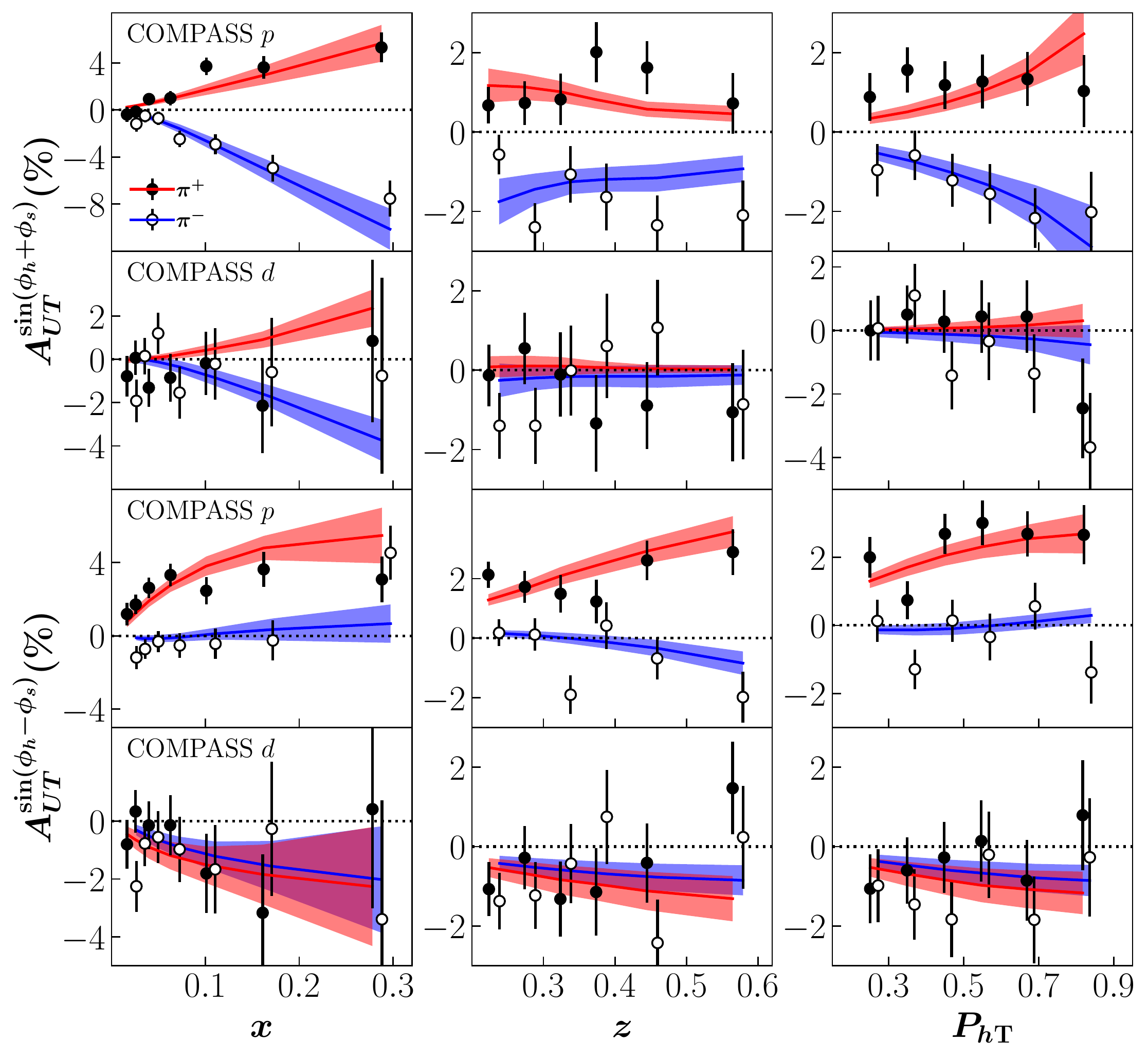}\vspace{-0.3cm}
\caption{Plot of the Collins $A_{UT}^{\sin(\phi_h+\phi_S)}$ (top two rows) Sivers $A_{UT}^{\sin(\phi_h-\phi_S)}$ (bottom two rows) asymmetries from COMPASS compared to our {\tt JAM3D-22}  global analysis.
} 
\label{f:sidis_COMPASS}
\end{figure}

\begin{figure}[h!]
\centering
\includegraphics[width=0.575\textwidth]{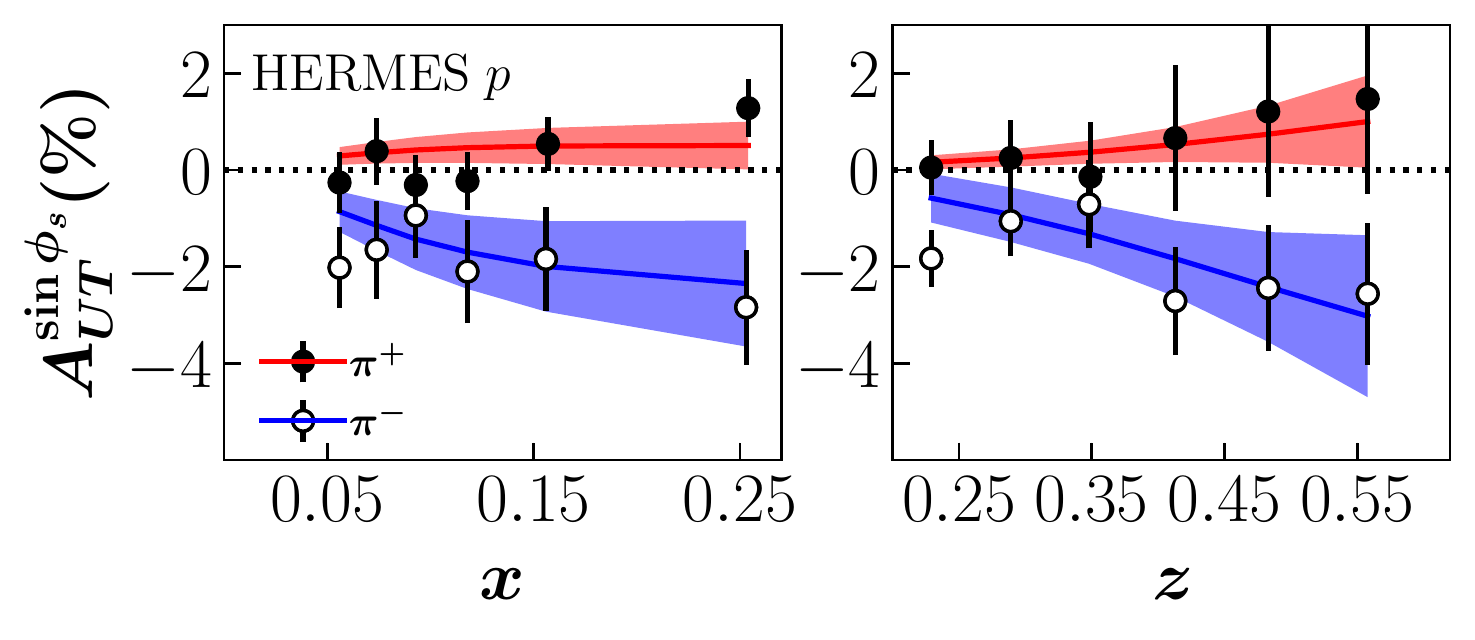}
\vspace{-0.3cm}
\caption{Plot of the $A_{UT}^{\sin\phi_S}$ asymmetry ($x$- and $z$- projections only) from HERMES compared to our {\tt JAM3D-22}  global analysis.}
\label{f:sidis_sinphiS}
\end{figure}

\begin{figure}[h!]
\centering
\includegraphics[width=0.42\textwidth]{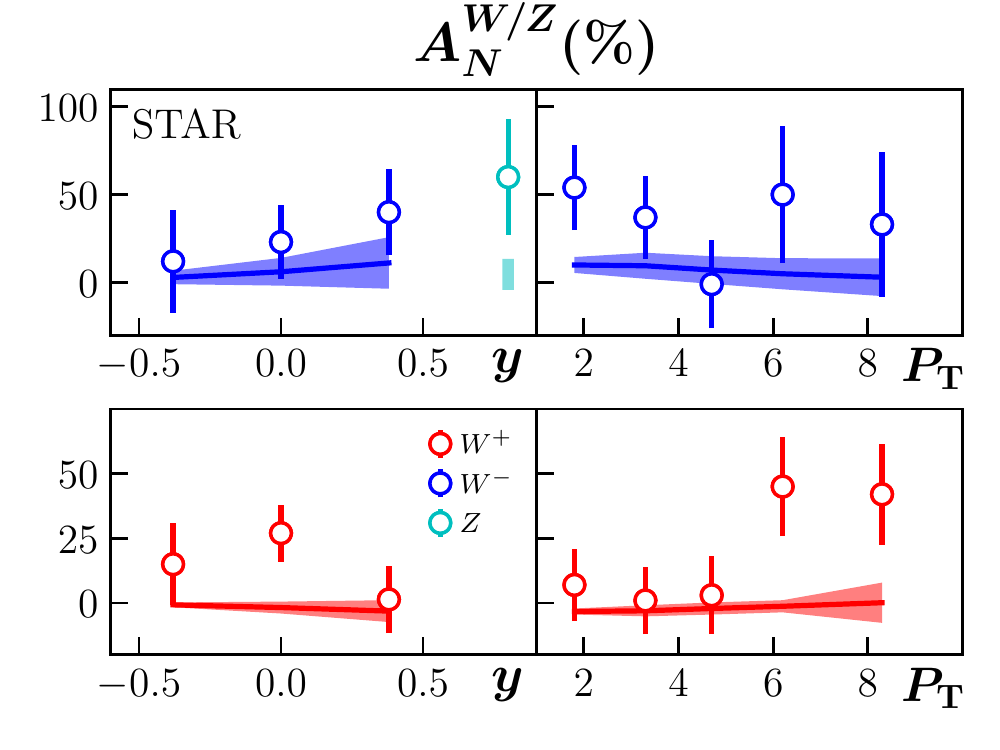}
\includegraphics[width=0.42\textwidth]{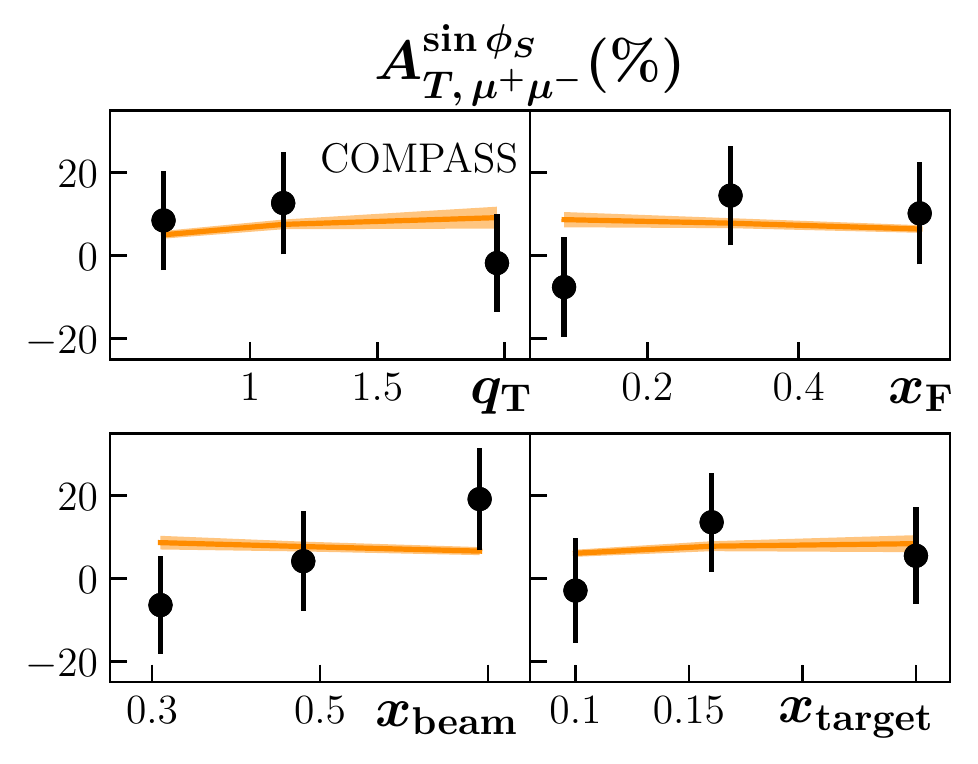}
\vspace{-0.3cm}
\caption{Plot of (left) the Sivers asymmetry in proton-proton DY $W/Z$ production from STAR ($A_N^{W/Z}$) and (right) pion-induced DY $\mu^+\mu^-$ production from COMPASS ($A_{T,\mu^+\mu^-}^{\sin\phi_S}$) compared to our {\tt JAM3D-22}  global analysis.
} 
\label{f:DY}
\end{figure}

\begin{figure}[h!]
\centering
\includegraphics[width=0.8\textwidth]{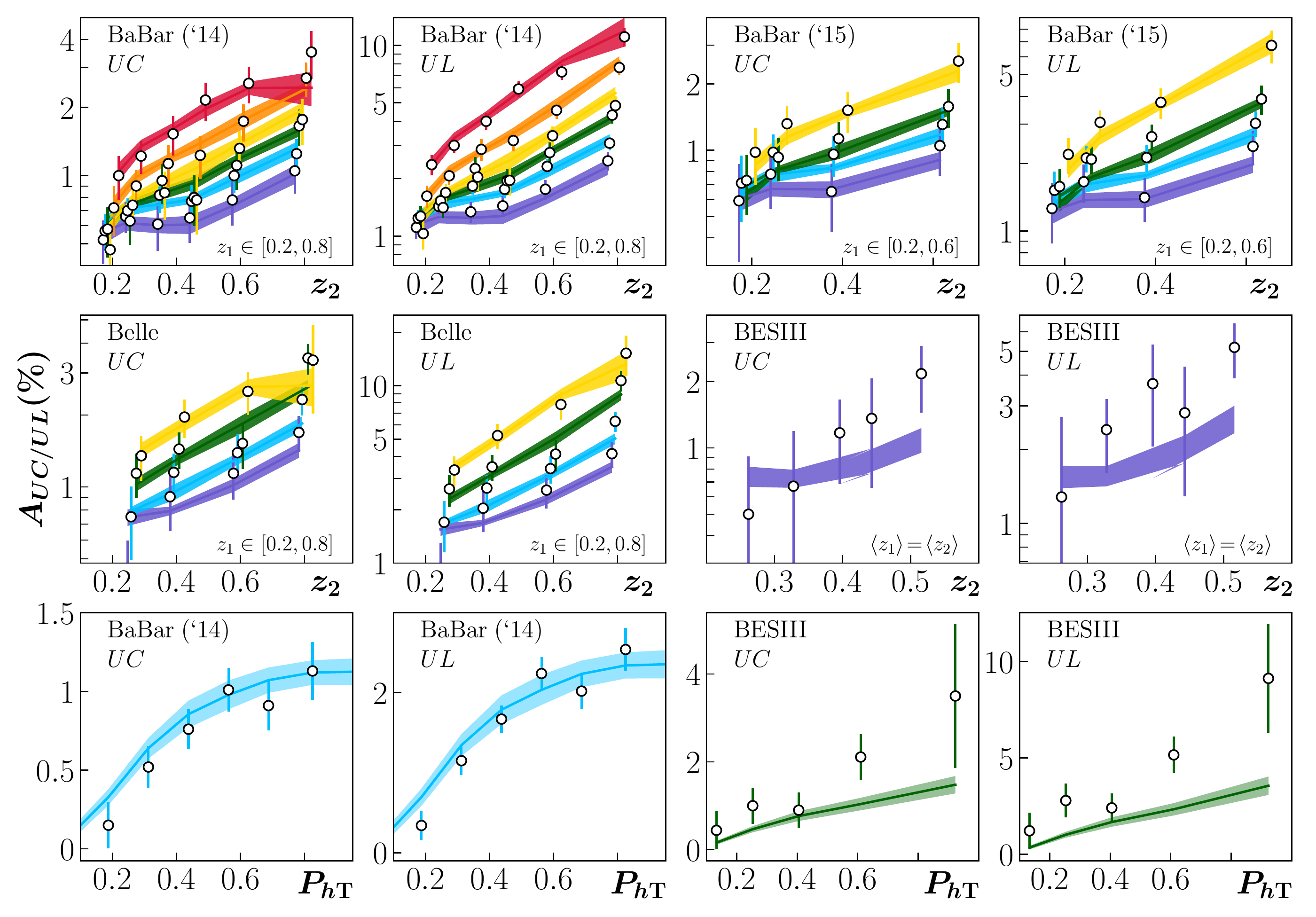}
\vspace{-0.3cm}
\caption{Plot of the Collins effect in SIA from Belle, BaBar, and BESIII for unlike-charged (UC) and unlike-like (UL) combinations compared to our {\tt JAM3D-22}  global analysis.
} 
\label{f:sia}
\end{figure}

\begin{figure}[h!]
\centering
\includegraphics[width=0.6\textwidth]{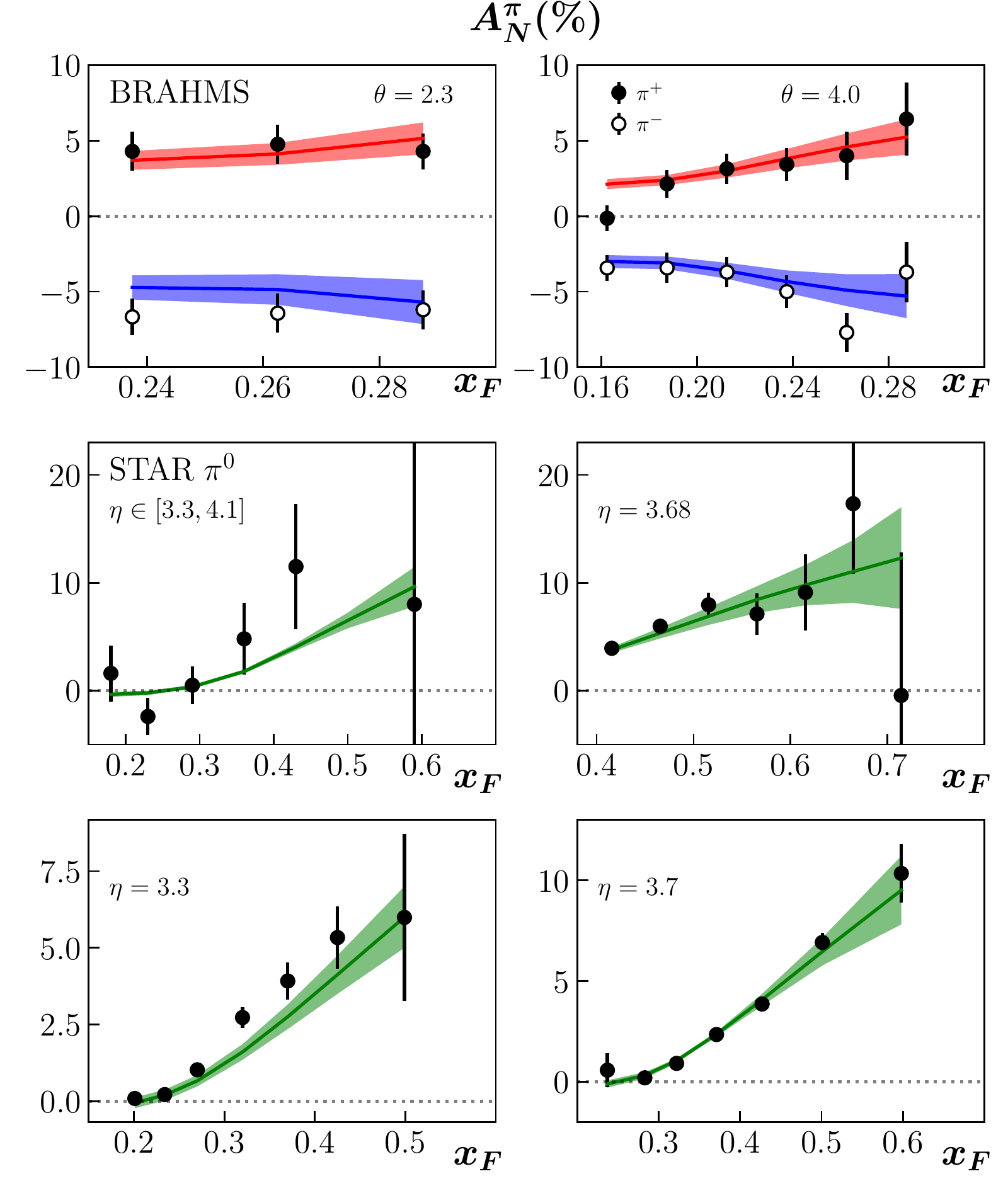}
\vspace{-0.3cm}
\caption{Plot of the proton-proton $A_N^\pi$ asymmetry from BRAHMS and STAR compared to our {\tt JAM3D-22}  global analysis.
} 
\label{f:an}
\end{figure}

%%%%%%%%%%%%%%%%%%%%%%%%%%%%%%%%
%%%%%%%%%%%%%%%%%%%%%%%%%%%%%%%%
In this appendix we provide plots comparing the theoretical results of our new global analysis of SSAs ({\tt JAM3D-22}) to the experimental data:~Collins and Sivers effects from HERMES (Figs.~\ref{f:sidis_HERMES_Collins}, \ref{f:sidis_HERMES_Sivers}); Collins and Sivers effects from COMPASS (Fig.~\ref{f:sidis_COMPASS}); $A_{UT}^{\sin\phi_S}$ from HERMES (Fig.~\ref{f:sidis_sinphiS}); Sivers effect in DY from STAR and COMPASS (Fig.~\ref{f:DY}); Collins effect in SIA from Belle, BaBar, and BESIII (Fig.~\ref{f:sia}); and $A_N^\pi$ from BRAHMS and STAR (Fig.~\ref{f:an}).
\vspace{-0.125cm}

\end{document}